\documentclass[fleqn,usenatbib]{mnras}
\usepackage{newtxtext,newtxmath}
\usepackage[T1]{fontenc}

\DeclareRobustCommand{\VAN}[3]{#2}
\let\VANthebibliography\thebibliography
\def\thebibliography{\DeclareRobustCommand{\VAN}[3]{##3}\VANthebibliography}

\usepackage{graphicx}	
\usepackage{amsmath}	
\usepackage{siunitx}
\usepackage[flushleft]{threeparttable}

\DeclareSIUnit\jansky{Jy}
\def\degr{$^\circ$}

\defcitealias{Chandra2012}{CF12}
\defcitealias{Granot2002}{GS02}

\title[A search for GRB afterglows with ASKAP]{A search for radio afterglows from gamma-ray bursts with the Australian Square Kilometre Array Pathfinder}
\author[J. K. Leung et al.]{James
K.\ Leung,$^{1,2,3}$\thanks{E-mail: jleu9465@uni.sydney.edu.au}
Tara Murphy,$^{1,3}$
Giancarlo Ghirlanda,${^4}$
David L.\ Kaplan,$^{5}$
\newauthor
Emil Lenc,$^{2}$
Dougal Dobie,$^{1,2,3,6}$
Julie Banfield,$^{2}$
Catherine Hale,$^{7}$
Aidan Hotan,$^{7}$
\newauthor
David McConnell,$^{2}$
Vanessa A. Moss,$^{2,1}$
Joshua Pritchard,$^{1,2,3}$
Wasim Raja,$^{2}$
\newauthor
Adam J. Stewart,$^{1}$ and
Matthew Whiting$^{2}$
\\
$^{1}$Sydney Institute for Astronomy, School of Physics, The University of Sydney, NSW 2006, Australia\\
$^{2}$CSIRO Astronomy and Space Science, PO Box 76, Epping, NSW 1710, Australia\\
$^{3}$ARC Centre of Excellence for Gravitational Wave Discovery (OzGrav), Hawthorn, VIC 3122, Australia\\
$^4$INAF -- Osservatorio Astronomico di Brera, via E. Bianchi 46, I--23807 Merate, Italy\\
$^5$Department of Physics, University of Wisconsin-Milwaukee, P.O. Box 413, Milwaukee, WI 53201, USA\\
$^6$Centre for Astrophysics and Supercomputing, Swinburne University of Technology, Hawthorn, VIC 3122, Australia\\
$^7$CSIRO Astronomy and Space Science, PO Box 1130, Bentley, WA 6102, Australia
}

\date{Accepted 2021 February 2. Received 2021 January 29; in original form 2020 December 1}

\pubyear{2021}

\begin{document}
\label{firstpage}
\pagerange{\pageref{firstpage}--\pageref{lastpage}}
\maketitle
\begin{abstract}
We present a search for radio afterglows from long gamma-ray bursts using the Australian Square Kilometre Array Pathfinder (ASKAP).
Our search used the Rapid ASKAP Continuum Survey, covering the entire celestial sphere south of declination $+41$\degr{}, and three epochs of the Variables and Slow Transients Pilot Survey (Phase 1), covering $\sim 5,000$ square degrees per epoch.
The observations we used from these surveys spanned a nine-month period from 2019 April 21 to 2020 January 11.
We crossmatched radio sources found in these surveys with 779 well-localised (to $\leq 15$\arcsec) long gamma-ray bursts occurring after 2004 and determined whether the associations were more likely afterglow- or host-related through the analysis of optical images.
In our search, we detected one radio afterglow candidate associated with GRB~171205A, a local low-luminosity gamma-ray burst with a supernova counterpart SN~2017iuk, in an ASKAP observation 511 days post-burst.
We confirmed this detection with further observations of the radio afterglow using the Australia Telescope Compact Array at 859 and 884 days post-burst.
Combining this data with archival data from early-time radio observations, we showed the evolution of the radio spectral energy distribution alone could reveal clear signatures of a wind-like circumburst medium for the burst.
Finally, we derived semi-analytical estimates for the microphysical shock parameters of the burst: electron power-law index $p = 2.84$, normalised wind-density parameter $A_* = 3$, fractional energy in electrons $\epsilon_{e} = 0.3$, and fractional energy in magnetic fields $\epsilon_{B} = 0.0002$.
\end{abstract}

\section{Introduction}
Long gamma-ray bursts (lGRBs) are produced by ultra-relativistic, narrowly collimated jets, originating from the core collapse of massive stars \citep{Woosley1993, MacFayden1999}.
Previous observations of core-collapse supernovae accompanying lGRBs have provided substantial evidence supporting this progenitor model; for example, GRB 980425/SN 1998bw \citep{Galama1998} and GRB 030329/SN 2003dh \citep{Stanek2003}.
In the standard fireball model, Fermi acceleration of electrons in the shock front resulting from jet interaction with the circumburst medium produces an afterglow visible across the entire electromagnetic spectrum \citep{Sari1998}.
In many cases, radio afterglow emission is detectable at late-time on the order of months to years post-burst, even when the jet has decelerated into the non-relativistic regime and its expansion has become quasi-spherical (\citealt{Frail2000}; \citealt{Chandra2012}, hereafter CF12).

The study of radio afterglows helps to constrain the microphysics and energetics (calorimetry) of burst events \citep{Frail2000, vanderHorst2008, Granot2014} and enables the reverse shock to be studied with greater observational latency compared with shorter wavelengths; for example, unlike the optical flash that occurs on the timescale of tens of seconds post-burst \citep{Akerlof1999, Sari1999}, the radio emission from the reverse shock persists for a few hours to a few days post-burst \citep{Kulkarni1999, Anderson2018}.
Radio afterglows observed with diffractive scintillation \citep{Goodman1997, Waxman1998} or Very Long Baseline Interferometry \citep[VLBI;][]{Granot1999, Taylor2004} techniques also provide measurements of the source size and jet expansion of burst events.
However, due to the limited availability of observing time, follow-up observations for radio afterglows have been biased towards known bursts that meet a specific set of selection criteria: for example, bursts with strong X--ray, optical \citep[\citetalias{Chandra2012};][]{Ghirlanda2013} or gravitational wave counterparts \citep{Nakar2011, Hallinan2017}; suspected dark bursts with optical counterparts obscured by dust that could still be detectable at radio wavelengths \citep{Djorgovski2001}; ultra-long GRBs with prompt emission lasting $\geq 1,000$ seconds \citep{Levan2014, Horesh2015}; and bursts at high redshifts that could probe Population-III stars \citep{Toma2011, Burlon2016}.

Unbiased radio transient surveys enable us to probe all events in the survey footprints for unusual or unexpected afterglow emission, which may otherwise be missed if not prioritised in traditional follow-up campaigns. For instance, using surveys to follow-up events could further our understanding of radio-quiet and -loud GRBs \citep{Hancock2013, LR2017, LR2019} as well as test for hypothesised GRB/supernova (GRB/SN) afterglow rebrightening effects, which are expected to occur on the order of decades post-burst when the supernova ejecta enters the Sedov-Taylor phase \citep{BD2015, Peters2019}. They could also enable us to monitor the late-time behaviour of well studied bursts with greater cadence, allowing for their burst parameters to be constrained independent of the burst geometry as the fireball transitions into the non-relativistic regime \citep[e.g.][]{Frail2000}.
Unlike studies using unbiased radio transient surveys for orphan afterglow searches, for example, \citet{Levinson2002} and \citet{Ghirlanda2014}, we limited our use of such surveys to the follow-up of on-axis bursts with a high-energy trigger.

The Australian Square Kilometre Array Pathfinder (ASKAP; Hotan et al. \emph{in press}) radio telescope operates at observing frequencies ranging from 700 to $1,800$\,MHz.
It has a $\sim 30$ square degrees field of view and is capable of reaching $\sim 1$\,\si{\milli\jansky\,beam^{-1}} RMS in 1\,min of integration time.
These capabilities facilitate observations for multi-epoch unbiased radio surveys, including the Rapid ASKAP Continuum Survey \citep[RACS;][]{McConnell2020} and the first phase of the pilot program for an ASKAP survey for Variables and Slow Transients \citep[VAST;][]{Murphy2013}.
We conducted a search for lGRB radio afterglows using RACS and three epochs of the VAST Pilot Survey (Phase 1; VAST-P1), with observations spanning a nine-month period from 2019 April 21 to 2020 January 11.

The structure of our paper is as follows.
We outline the observations and data quality for RACS and each epoch of VAST-P1 used in our search in \textsection\ref{sec:datasets}.
In \textsection\ref{sec:rates}, we outline the expected detection rates of radio afterglows in our search.
In \textsection\ref{sec:detectability}, we explore the impact of host galaxy contamination on the detectability of radio afterglows in our search (and more generally, in a search for explosive transients).
In \textsection\ref{sec:tsearch}, we describe our search methodology and how we distinguished host galaxy emission from afterglow emission.
In \textsection\ref{sec:results}, we describe our detection and follow-up of GRB 171205A, an interesting source we identified in \textsection\ref{sec:tsearch}.
In \textsection\ref{sec:discussion}, we discuss the physical interpretation of our GRB 171205A radio observations, the consideration of alternative search strategies, and the expected GRB yields for comparison surveys (both current and future).
Finally, we summarise the conclusions from our radio afterglow search in \textsection\ref{sec:conclusions}.

We adopt a flat $\Lambda$-CDM cosmology with $H_0=~67.8$\,km\,s$^{-1}$\,Mpc$^{-1}$, $\Omega_{\text{M}} = 0.308$ and $\Omega_{\Lambda} = 0.692$ throughout this paper \citep{Planck2016}.

\section{Observations and Data Analysis}
\label{sec:datasets}

\begin{figure}
	\includegraphics[width=\columnwidth,clip,trim={0.2cm 2.5cm 0.3cm 2.5cm}]{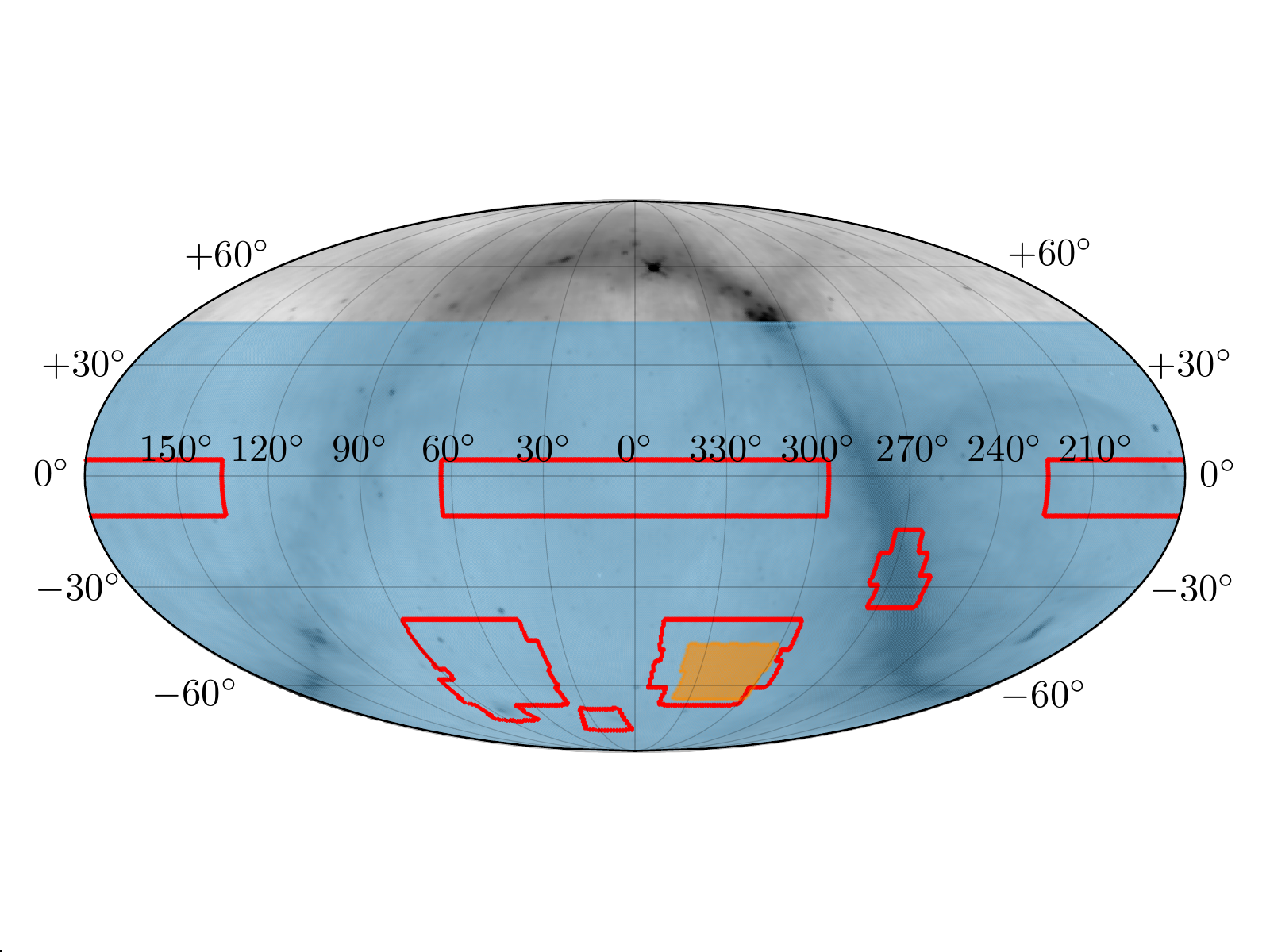}
    \caption{The VAST-P1 footprint is outlined in red and consists of 113 tiles. The RACS footprint shaded in blue extends between $-$90\degr $\leq\delta<+41$\degr. The EMU Pilot Survey shaded in orange consists of 10 tiles overlapping with the VAST-P1 and RACS surveys. The sky map is plotted with J2000 equatorial coordinates in the Mollweide projection and the background diffuse Galactic emission at 887.5\,MHz (gray) is modelled from \citet{Zheng2017}.}
    \label{fig:vastskymap}
\end{figure}

\begin{table*}
	\centering
	\caption{Epochs from the ASKAP surveys, RACS and VAST-P1, were used in our GRB radio afterglow search. Columns 1 through 8 show the epoch number, radio survey used for each epoch, start date of each epoch, solid angle and number of tiles covered by each epoch, median RMS noise of observations in each epoch, median right ascension offset from ICRF2 source positions for observations in each epoch, median declination offset from ICRF2 source positions for observations in each epoch, and the flux-density ratio of sources in each epoch compared against sources in SUMSS. Observations for all epochs were conducted at the central frequency of 887.5\,MHz with a bandwidth of 288\,MHz and an angular resolution of $\sim 15\arcsec$.}
	\label{tab:datasummary}
	\begin{threeparttable}
	\begin{tabular}{cccccccccc} 
		\hline
		\hline
		Epoch \# & 
		Survey &
		Start Date (UT) &
		Sky Coverage & 
		\begin{tabular}[c]{@{}c@{}}RMS\\ (\si{\milli\jansky\,beam^{-1}})\end{tabular} & 
		\begin{tabular}[c]{@{}c@{}}RA Offset\\(arcsec)\end{tabular} & 
		\begin{tabular}[c]{@{}c@{}}Dec Offset\\(arcsec)\end{tabular} & 
		$S_{\text{ASKAP}}/S_{\text{SUMSS}}$\\
		\hline
		
		0 & RACS & 2019 Apr 21 & $-$90\degr $\leq \delta<+$41\degr, 903 tiles & 0.29 & $-0.6 \pm 0.8$ & $+0.2 \pm 0.8$ & $1.03 \pm 0.29$ \\
		
		1 & VAST-P1 & 2019 Aug 27 & $\sim 4$,980 sq. deg, 113 tiles & 0.29 & $-0.1 \pm 0.6$ & $+0.5 \pm 0.9$ & $1.01 \pm 0.25$ \\
		
		2 & VAST-P1 & 2019 Oct 28 & $\sim 4$,760 sq. deg, 108 tiles & 0.28 & $+0.2 \pm 0.6$ & $+0.2 \pm 1.0$ & $1.05 \pm 0.30$ \\
		
		8 & VAST-P1 & 2020 Jan 11 & $\sim 4$,940 sq. deg, 112 tiles & 0.26 & $-0.4 \pm 0.6$ & $+1.0 \pm 1.2$ & $1.06 \pm 0.19$ \\

		\hline
	\end{tabular}
	\begin{tablenotes}
      \small
      \item \textit{Note} -- Epoch 8 is the third full epoch of the VAST-P1 survey. Several partial epochs between Epoch 2 and 8 were observed, but not used in our search. See O'Brien et al. (\emph{in prep.}) for more details.
    \end{tablenotes}
	\end{threeparttable}
\end{table*}

The observations and quality metrics associated with the two ASKAP surveys used in our search are described in this section. Observations in both surveys were conducted at a central frequency of 887.5\,MHz with a bandwidth of 288\,MHz and a typical angular resolution of $\sim 15$\arcsec.

\subsection{Rapid ASKAP Continuum Survey}
RACS is a large-area survey consisting of 903 tiles covering the entire radio sky south of declination $+41$\degr.
Each tile had an integration time of $\sim 15$\,min and the resulting images had a median RMS noise of $\sim 0.29$\,\si{\milli\jansky\,beam^{-1}}.
The RACS observations used for our search were conducted between 2019 April 21 and 2019 November 22.
The full set of RACS observations include some tiles that were reobserved, with the final observations occurring on 2020 June 21.
These subsequent observations were not used in our search, but were used in a candidate radio afterglow follow-up discussed in \textsection\ref{sec:results}.
In this paper, RACSe1 refers to the initial observation of a tile and RACSe2 refers to any subsequent observation of the same tile.

We used a pre-release version of the RACS data for our radio afterglow search.
As a result, this version of the RACS data did not implement some of the improvements to the data quality featured in the published data products.
These improvements included, for example, holography corrections for the direction dependent flux-density scale and techniques to account for variations of the point-spread-function (PSF) size and shape across the ASKAP field of view.
A more detailed discussion of these issues and their corresponding improvements are found in \citet{McConnell2020} and Hale et al. (\emph{in prep.}).
Please refer to these two papers for comprehensive details on the survey design, calibration process, data reduction strategy as well as the data products associated with the published images and source catalogues.

We performed an independent quality control check on the astrometric accuracy and flux-density scale to the aforementioned papers since the version of the data used for our search differs from the published data products.
For our quality control analysis specifically, we used the {\sc Selavy} \citep{Whiting2012} source finder with default settings to extract sources from the RACS images and filtered for sources that were (a) isolated ($\geq 150$\arcsec\ from nearest neighbour), (b) with non-extended PSF (major axis length $\leq 50$\arcsec\ with a major-to-minor axis ratio $\leq2$), and (c) had a signal-to-noise ratio $\geq 10$.
The reason we chose a non-extended PSF criterion over a conventional point-source criterion (e.g. peak-to-integrated flux-density ratio) was due to aforementioned PSF variations affecting measurements of the integrated flux-density.
This set of criteria ensured only isolated sources with reasonably compact PSFs and high signal-to-noise were used for the quality control analysis in this section; we did not apply these criteria to filter sources for our radio afterglow searches in subsequent sections of this paper.

Using this subset of sources, we determined the astrometric offset of RACS sources from the position of sources from the second realisation of the International Celestial Reference Frame \citep[ICRF2;][]{Fey2015}; the offset is $-0\farcs6\pm 0\farcs8$ in right ascension and $+0\farcs2\pm 0\farcs8$ in declination.
We calibrated the flux-density scale using the same subset of sources against the Sydney University Molonglo Sky Survey \citep[SUMSS;][]{Mauch2003}, which was conducted at a comparable observing frequency of 843\,MHz.
This yielded a median flux-density ratio of $S_{\text{RACS}}/S_{\text{SUMSS}} = 1.03 \pm 0.29$, assuming a spectral index $\alpha$ of $-0.8$ \citep{Condon1992}, defined as $S_\nu \propto \nu^{\alpha}$.
We have thus independently verified the reliability of the astrometry and flux-density scale for our early data products used in our scientific analysis.

\subsection{VAST Pilot Survey (Phase 1)}
Table \ref{tab:datasummary} summarises the datasets from the three VAST-P1 epochs used in our search.
VAST-P1 covers 113 tiles using the same tiling footprint as RACS; see Figure \ref{fig:vastskymap} for the sky coverage of both surveys.
The integration time per tile per epoch was $\sim 12$\,min and the resulting images for each epoch had a median RMS noise varying from $0.26$ to $0.29$\,\si{\milli\jansky\,beam^{-1}}.
Some epochs have a few ($\leq 5$) tiles missing due to various issues (e.g. solar interference, observation failures, etc.) discussed in O'Brien et al. (\emph{in prep.}).
All observations were processed using standard procedures in the {\sc ASKAPsoft} package \citep{Cornwell2011, Guzman2019} and sources were extracted with the {\sc Selavy} \citep{Whiting2012} source finder.
The full details of the observing strategy as well as the calibration, reduction and source extraction procedures are described in O'Brien et al. (\emph{in prep.}).

Similar to before, we used a pre-release version of the VAST-P1 data, and hence, performed an independent quality control check.
Applying the same procedure from the RACS quality control analysis, we found the astrometry of each VAST-P1 epoch to be typically accurate to $<1$\arcsec\ in both right ascension and declination, while the flux-density scale was typically calibrated to within 6 per cent from that of SUMSS with a RMS scatter $\leq 30$ per cent.
The astrometric accuracy and flux-density scale for each epoch are listed in Table~\ref{tab:datasummary}.
We have thus demonstrated that the data quality for VAST-P1 early data products was comparable to the RACS early data products; both were sufficient for the analysis described in subsequent sections, involving analyses of positional offsets and source variability.

\section{Expected Afterglow Detection Rate}
\label{sec:rates}
We constrained the expected afterglow detection rate for on-axis afterglows in our search using archival data of previous observations.
In this work, we only consider the lGRB afterglow detections as short gamma-ray burst (sGRB) afterglow detections require flux-density sensitivities well below the thresholds of the surveys we use in this work \citep[e.g.][]{Fong2015}.
To determine the expected lGRB afterglow detection rate, we used the \citetalias{Chandra2012} catalogue, consisting of 304 on-axis lGRB afterglows observed by radio telescopes over a 14 year span from 1997 to 2011.
Radio emission was detected from 95 of these lGRBs, and of these, only 64 had known redshifts and fitted maximum flux-density measurements.
In this paper, `maximum flux density' refers to the flux density of a radio afterglow at the light curve peak for the observing frequency of interest.
Some other papers in the literature (e.g. \citetalias{Chandra2012}) refer to this as the `peak flux density,' but we choose to use different terminology in our paper to avoid any confusion with standard radio astronomy terminology.

\begin{figure*}
	\includegraphics[width=0.85\linewidth,clip,trim={0.32cm 0.5cm 0.35cm 0.4cm}]{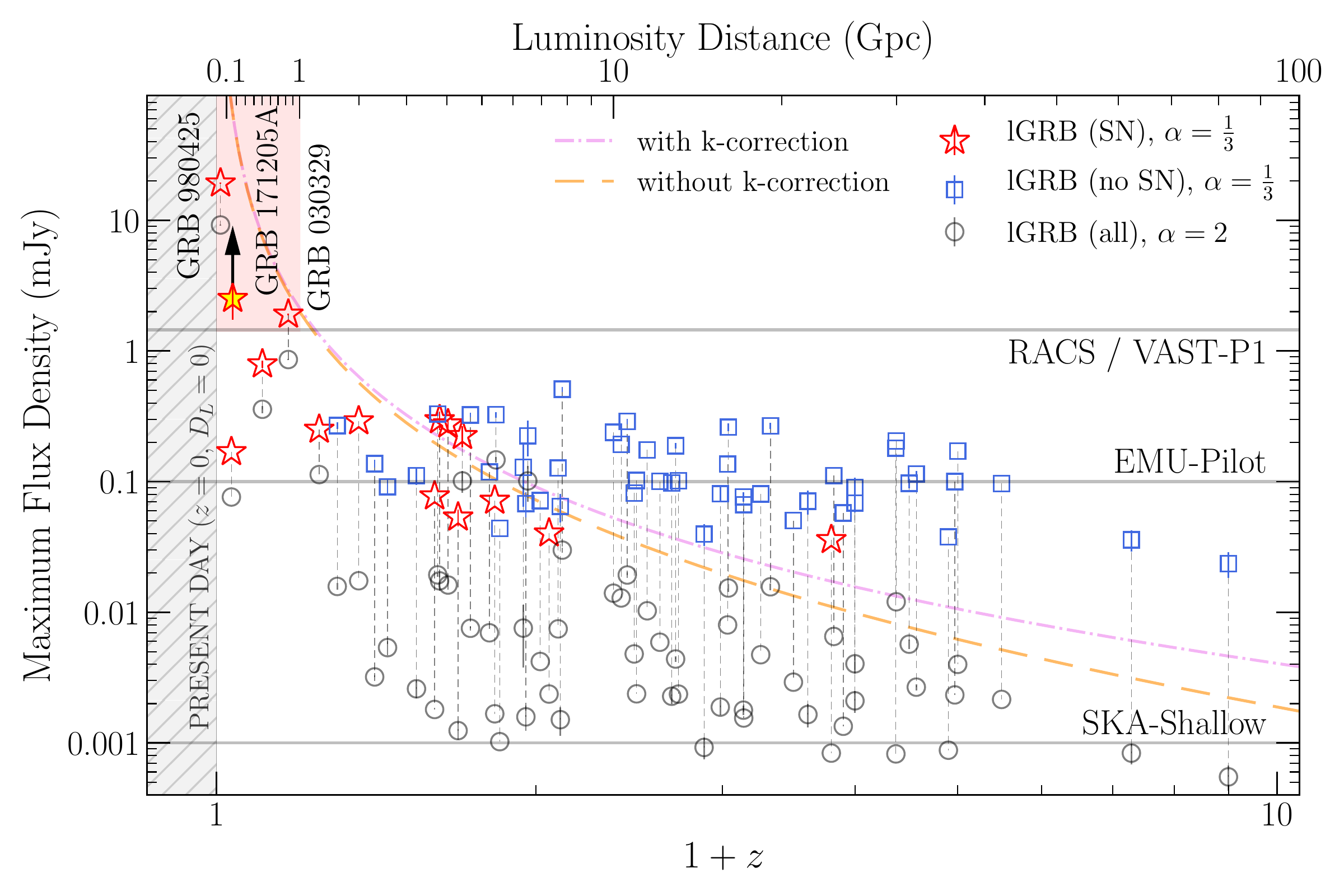}
    \caption{
    The fitted maximum flux density scaled to 887.5\,MHz as a function of $1+z$ (bottom axis) and luminosity distance $D_{L}$ (top axis) for 64 radio afterglows featured in \citetalias{Chandra2012}.
    The maximum flux density refers to the flux density of a radio afterglow at the light curve peak.
    Each burst is scaled with two different spectral indices, $\alpha = \frac{1}{3}$ (self-absorption is not significant) and $\alpha = 2$ (self-absorption is significant), providing upper and lower bounds, respectively.
    For the $\alpha = \frac{1}{3}$ scenario, the 14 afterglows with supernova association are plotted with red star markers, while those without are plotted with blue square markers.
    For the $\alpha = 2$ scenario, all afterglows are plotted with gray circular markers.
    For each burst, the scaled data points using the two different spectral indices are joined by gray dashed lines.
    The error bars represent uncertainties in fitting for the maximum flux density, not measurement uncertainties; data points without error bars are measured values, not fitted values; see \citetalias{Chandra2012} for details.
    The expected relations between the maximum flux density and redshift, assuming a typical spectral luminosity of $L_\nu = 2 \times 10^{30}$\,erg\,s$^{-1}$\,Hz$^{-1}$, are shown for two separate cases: (i) the relation governed by the typical inverse square decrease of the maximum flux density with luminosity distance is shown with the orange dashed line, (ii) the relation accounting for negative $k$-correction effects leading to the plateauing effect at cosmological distances is shown with the purple dash-dotted line.
    For a more detailed discussion of the scaling and the dependence of maximum flux density on redshift, see \textsection\ref{sec:rates}.
    The gray horizontal lines show the 5$\sigma$-limit of different radio surveys considered in this work.
    The region shaded in red above $1.50$\,\si{\milli\jansky} and below $z = 0.2$ ($D_{L} = 1$\,Gpc) represents the parameter space our search is most sensitive to.
    The lower limit for the maximum flux density of our GRB 171205A afterglow candidate (yellow star marker) is given by its flux density as measured from RACS.
    }
    \label{fig:onaxisrates}
\end{figure*}

Figure \ref{fig:onaxisrates} shows the scaled maximum flux densities against redshift and luminosity distance for radio-detected afterglows from the sample as well as the $5\sigma$ flux-density threshold for various surveys considered in this work.
Note that while the threshold of the Evolutionary Map of the Universe \citep[EMU;][]{Norris2011} Pilot Survey is lower than the RACS/VAST-P1 threshold, it was unsuitable for our search due to the limited sky coverage (see Figure \ref{fig:vastskymap}).
We scaled the fitted maximum flux density of the lowest frequency observation for each GRB in \citetalias{Chandra2012} to 887.5\,MHz, which is the central frequency of RACS and VAST-P1 observations.
Two different spectral indices were applied for the scaling, $\alpha = 1/3$ and $\alpha = 2$, corresponding to different scenarios for the self-absorption frequency, $\nu_{\text{sa}}$.
In both these scenarios, we considered the typical afterglow spectrum in the slow cooling regime as described in \citet[][hereafter GS02]{Granot2002}.
In the $\alpha = 1/3$ scenario, our observing frequency is assumed to be above $\nu_{\text{sa}}$ and below the synchrotron frequency corresponding to the peak, $\nu_{\text{m}}$, from which we scaled; this corresponds to power-law segment D in \citetalias{Granot2002}.
In the $\alpha = 2$ scenario, our observing frequency is assumed to be below $\nu_{\text{sa}}$, corresponding to power-law segment B in \citetalias{Granot2002}.
In this scenario, the maximum flux density at the frequency we scaled from could either be due to the passage of $\nu_{\text{m}}$ (in an ISM environment) or $\nu_{\text{sa}}$ (in a wind environment); in both cases, the appropriate spectral scaling is still $\alpha = 2$.
Predicting the temporal evolution of the flux density at $\nu_{\text{m}}$ as it moves towards our observing frequency requires further modelling for each burst; we consider this beyond the scope of our work and argue that the small difference between the frequency we scaled from (typically 1.43 or 8.46\,GHz) and our observing frequency would not result in substantial changes to the flux density at $\nu_{\text{m}}$ as it moves towards our observing frequency.
We therefore only used the two spectral indices to provide a range on the expected maximum flux density for each burst when scaled to 887.5\,MHz and noted that this range should be shifted upwards slightly if the temporal evolution of the synchrotron peak was also factored in.

Considering the $\alpha = 1/3$ spectral scaling, only afterglows from GRB 980425 and GRB 030329 would be detectable in RACS and VAST-P1 at the 5$\sigma$-level during the 14 year span covered by \citetalias{Chandra2012}. From archival data, the duration, or length of time the radio afterglow would be detectable, for GRB 980425 and GRB 030329, would be approximately 250 \citep{Li1999} and 150 days \citep{Frail2005}, respectively. 
The implied fractional time a radio afterglow from a known GRB is detectable in the sky at the RACS/VAST-P1 flux-density sensitivity is 7.8 per cent.
We calculated the probability of detecting a radio afterglow in our search to be 7.2 per cent after factoring in the sky coverage of RACS and VAST-P1 as well as the cadence of our observations.
These calculations should be considered as lower limits since less than 20 per cent of known GRBs have been observed at radio frequencies and early-time radio detections from the reverse shock mechanism were not considered in \citetalias{Chandra2012}.
Considering the $\alpha = 2$ spectral scaling, only GRB 980425 would be detectable in RACS and VAST-P1, implying a probability for detecting a radio afterglow in our search to be 3.7 per cent.
For comparison, the number of on-axis radio afterglow detections predicted in \citet{Ghirlanda2013} and \citet{Metzger2015} is of order unity when their predictions for similar surveys are scaled to our search parameters.

Despite the low probability of detecting a typical radio afterglow, the detections of GRB 980425 and GRB 030329 implied that finding a subset of local, radio-bright afterglows in our search was possible.
These two GRBs, associated with SN 1998bw \citep{Galama1998,Kulkarni1998} and SN 2003dh \citep{Hjorth2003,Stanek2003} respectively, suggest this subset of GRBs at redshifts $z < 0.2$ (or luminosity distances $D_{L} < 1$\,Gpc) are often sufficiently local for accompanying broad-lined Type Ic (Ic-BL) supernovae to also be detected.
This is a class of supernovae with spectral lines broadened from quasi-relativistic photospheric velocities reaching $\sim 30,000$\,km\,s$^{-1}$ \citep{Modjaz2016} and with kinetic energies reaching $\sim 10^{52}$\,erg when accompanied by lGRBs \citep{Mazzali2014}.
At more cosmological distances probed by deeper searches, the identification of accompanying supernovae becomes more difficult, but the dependence of the afterglow maximum flux density on redshift becomes weaker due to the negative $k$-correction effect \citep{Ciardi2000, Frail2006}.
This is represented in Figure \ref{fig:onaxisrates}: the conventional inverse square luminosity distance decrease (orange dashed line) follows the relation $L_\nu = 4\pi S_\nu D_{L}^2/(1+z)$, where $L_\nu$ is the spectral luminosity at the observing frequency, $S_\nu$ is the flux density, $D_{L}$ is the luminosity distance and $z$ is the redshift; the negative $k$-correction factor $(1+z)^{\,\beta-\alpha}$ multiplied to the aforementioned relation results in a flatter flux-density dependence on redshift (purple dash-dotted line); these lines assume a typical spectral luminosity of $L_\nu = 2 \times 10^{30}$\,erg\,s$^{-1}$\,Hz$^{-1}$ \citepalias{Chandra2012} and a post jet break light curve that is optically thin and flat \citep[i.e. spectral index $\alpha = 1/3$, temporal index $\beta = 0$, where $S \propto \nu^{\,\alpha} t^{\,\beta}$;][]{Frail2006}\footnote{The assumption of $\alpha=1/3$ holds only for the blue/red (not gray) points in Figure \ref{fig:onaxisrates} while the $\beta = 0$ (flat light curve) carries the assumption of a wind-like circumburst medium. Assuming $\alpha=2$ (e.g. gray points) and/or an ISM-like circumburst medium with different temporal scalings will therefore result in a different $k$-correction factor (and associated lines).}.
Therefore, the redshift limitations impacting our search would be less severe at the flux-density sensitivities achieved by the deeper EMU survey, and to a larger extent, the Square Kilometre Array Shallow \citep[SKA-Shallow; e.g.][]{Fender2015} survey in the future.

\section{Detectability of Afterglows}
\label{sec:detectability}
The detectability of afterglows in our search depended on the flux-density sensitivity of our surveys and the degree of host galaxy contamination present.
Only radio afterglows that reached our 5$\sigma$-sensitivity threshold of $\sim 1.50$\,\si{\milli\jansky\,beam^{-1}} during our observations could be detectable in our search.
We expected our detections to be close to this 5$\sigma$-threshold since the majority of previous radio afterglow detections peaked below this threshold level (see \textsection\ref{sec:rates}). 

\begin{figure}
	\includegraphics[width=\columnwidth,clip,trim={0.075cm 0.1cm 0.1cm 0cm}]{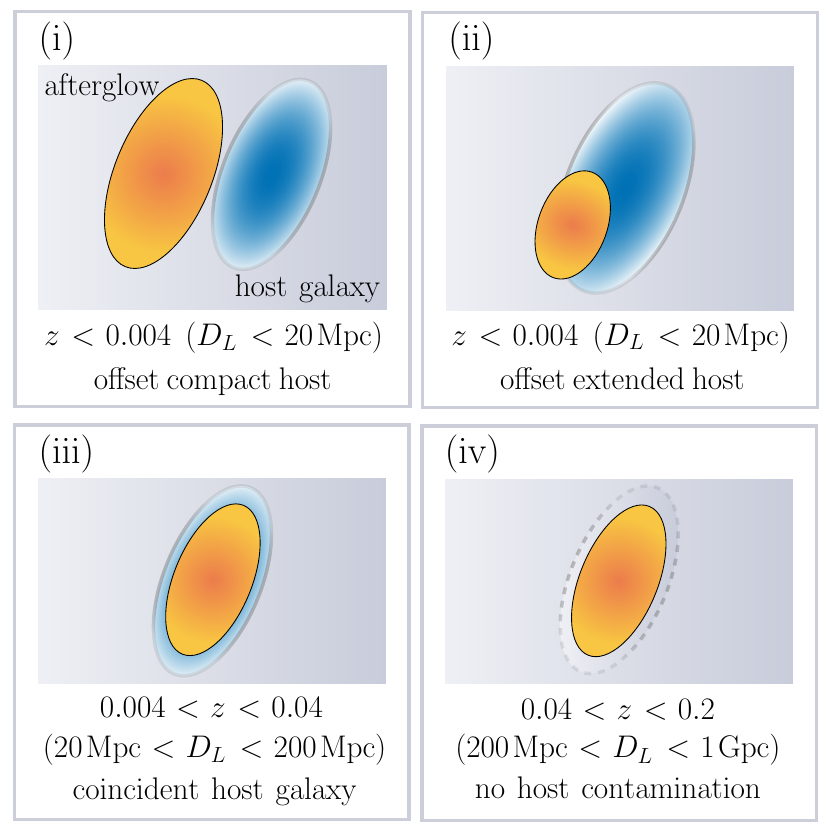}
    \caption{
    Four different scenarios showing how the detectability of radio afterglows may be contaminated by the radio emission of the host galaxies.
    The typical redshift (and luminosity distance) range for each scenario is given assuming: (a) the typical physical offset of an lGRB afterglow from its host to be $\sim 1.3$\,kpc, (b) the typical star formation rate of a host galaxy to be 1\,$M_\odot\,\text{yr}^{-1}$, and (c) the observations are made using a radio telescope with $\sim 15$\arcsec\ angular resolution.
    A detailed discussion of these assumptions can be found in \textsection\ref{sec:detectability}.
    }
    \label{fig:agdet_schematic}
\end{figure}

A radio afterglow emitting above the threshold still needed to be distinguishable from its host galaxy.
The radio emission of local galaxies at redshifts $z\ll 1$ follow the relationship:
\begin{equation}
    S_{\nu,\,\text{gal}} = 0.3\,\text{mJy}\,\bigg(\frac{\text{SFR}}{M_\odot\ \text{yr}^{-1}}\bigg)
    \,\nu^{-0.8}_{\text{GHz}}\,
    D^{-2}_{L,27}\,,
	\label{eq:hgflux}
\end{equation}
where SFR is the star formation rate in units of $M_\odot\ \text{yr}^{-1}$, $\nu_{\text{GHz}}$ is the observing frequency in GHz, $D_{L}~=~10^{27}D_{L,27}$\,cm is the luminosity distance to the host galaxy and the applied spectral index of $-0.8$ characterises the synchrotron emission of radio galaxies at gigahertz frequencies \citep{Condon1992, Carilli1999, Metzger2015}. The problem of host galaxy contamination was considered for each of the four scenarios below, depending on the radio afterglow localisation with respect to its host galaxy. The four scenarios we discuss are illustrated in Figure \ref{fig:agdet_schematic}.

\begin{enumerate}
	\item \textit{The afterglow is offset and resolvable from its host galaxy.} Host galaxies in this scenario are compact and have angular separations from their afterglows that are at least as large as the angular resolution of our observations ($\geq 15$\arcsec\ for RACS/VAST-P1). Afterglows in this scenario would be detectable if their peak flux densities are above our 5$\sigma$-sensitivity threshold of $\sim 1.50$\,\si{\milli\jansky\,beam^{-1}}.
	\item \textit{The afterglow is offset but not clearly resolved from its host galaxy.} Host galaxies that are very local \citep[e.g. $z < 0.03$ for radio galaxy samples in][]{vanVelzen2012} may have extended emission at gigahertz frequencies. An afterglow offset from its host galaxy centre, even by more than 15\arcsec, may be contaminated by this emission and our 5$\sigma$-sensitivity threshold criteria would not be enough to determine the detectability of the afterglow. With a RMS scatter in the flux-density scale of $\sim 0.3$ in RACS/VAST-P1 (see Table \ref{tab:datasummary}), a 5$\sigma$-detection requires the afterglow to reach flux densities $\geq 2.5$ times the underlying host emission as measured from previous epochs.
	If there exists a reference epoch prior to the afterglow detection, the host emission could be modelled and subtracted to varying degrees of fidelity; in this case, the detectability would only depend on the 5$\sigma$-sensitivity threshold criteria.
	\item \textit{The afterglow is spatially coincident with its host galaxy.} Host galaxies in this scenario are compact and have angular separations from their afterglows that are smaller than half the angular resolution of our observations ($\leq 8$\arcsec\ for RACS/VAST-P1). Similar to scenario (ii), any afterglow that is spatially coincident with its host galaxy would only be detectable in our search with 5$\sigma$-confidence if it reaches flux densities $\geq 2.5$ times the flux density of its host galaxy in the epochs prior to detection unless its host emission could be modelled from a reference epoch and subtracted.
	\item \textit{The afterglow is found without any detectable host galaxy emission associated with it.} Host galaxies of afterglows in this scenario are too faint to be detected in any epochs of our observations. These afterglows will not be contaminated by host galaxy emission and would be detectable if their peak flux densities are above our 5$\sigma$-sensitivity threshold of $\sim 1.50$\,\si{\milli\jansky\,beam^{-1}}.
\end{enumerate}

Physical offsets for lGRBs from their host galaxies range from 0.075 to 14\,kpc with a median offset of $\sim 1.3$\,kpc \citep{Bloom2002, Blanchard2016}.
Assuming the median (maximum) lGRB physical offset, an afterglow would be detected with a 15\arcsec\ angular offset from its host galaxy if it is at a redshift of 0.004 (0.046), or equivalently, a luminosity distance of 20\,Mpc (200\,Mpc). This implies that finding afterglows in scenarios (i) and (ii) with appreciable offsets from their hosts using ASKAP surveys is unlikely since all GRBs found in our sample in \textsection\ref{sec:rates} were located at $z > 0.004$ ($D_L > 20$\,Mpc). For instance, our lowest redshift sample, GRB~980425, at $z=0.008$ ($D_L = 40$\,Mpc) had a measured angular offset of $12\farcs55 \pm 0\farcs052$ \citep{Bloom2002}, which is below the angular resolution of our observations. This suggests radio afterglows found appreciably offset from their hosts would likely be associated with an extremely local burst $z < 0.004$ ($D_L < 20$\,Mpc).

The star formation rate of lGRB host galaxies range from 0.2 to 50\,$M_\odot\,\text{yr}^{-1}$ \citep{Berger2009} with a median of about 2\,$M_\odot\,\text{yr}^{-1}$ \citep{Christensen2004}.
Assuming a host galaxy star formation rate of 1\,$M_\odot\,\text{yr}^{-1}$, we argue using Equation \ref{eq:hgflux} that host galaxy contamination only becomes significant, i.e. $S_{\nu,\,\text{gal}} \geq 1$\,\si{\milli\jansky} ($\sim 3\sigma$), at $z < 0.04$ ($D_L < 200$\,Mpc).
For galaxies with a more active star formation rate of 10\,$M_\odot\,\text{yr}^{-1}$, the host galaxy contamination would be significant for bursts detected at $z < 0.12$ ($D_L < 580$\,Mpc).
Since we expected the brightest radio afterglows to be detectable at flux densities $\geq 1.50$\,\si{\milli\jansky} up to $z=0.2$ ($D_L = 1$\,Gpc), we expected more afterglow detections to fall under scenario (iv) than (iii) as host galaxy contamination was more likely to not be significant.

Consideration of other transients with different offset and luminosity distributions could have different conclusions.
For example, typical radio supernovae would be more likely found in scenarios (i) or (ii) because they have larger typical offsets and are detected more abundantly at luminosity distances $D_L < 20$\,Mpc due to their lower typical luminosities \citep{Wang1997}.
Likewise, radio telescopes with higher angular resolution are also more likely to result in detections corresponding to scenarios (i) or (ii).
Radio afterglows detectable in our search needed to be distinguishable from other transient phenomena or disentangled from any coincident background sources.
This required further multi-wavelength analysis and telescope time as demonstrated later in \textsection\ref{sec:results}.

\section{Search Methodology}
\label{sec:tsearch}
We compiled a comprehensive catalogue of lGRBs detected in the post-\textit{Swift} era, specifically between 2004 December and 2020 January (inclusive).
This consisted of $3,005$ bursts primarily detected by the International Gamma-Ray Astrophysics Laboratory \citep[INTEGRAL;][]{Mereghetti2003}, the \textit{Swift} Burst Alert Telescope \citep[BAT;][]{Lien2016} and the \textit{Fermi} Large Area Telescope \citep[LAT;][]{Ajello2019} and Gamma-ray Burst Monitor \citep[GBM;][]{vonKienlin2020}.
These missions typically localised the bursts to a few arcminutes, with the exception of the \textit{Fermi}-GBM, which had a cruder typical localisation radius of a few degrees.
A better localisation to a few arcseconds is often achieved with the automatic search for X--ray and optical afterglows following a trigger from the aforementioned missions with the \textit{Swift} X--ray Telescope \citep[XRT;][]{Evans2009} and \textit{Swift} Ultraviolet/Optical Telescope \citep[UVOT;][]{Roming2017}, respectively.
For our search method and analysis, we required localisation to the level of arcseconds so we only considered lGRBs detected by \textit{Swift}; the wider catalogue we compiled consisting of bursts from other missions was then used as a reference catalogue to assess our search strategy and completeness against in \textsection\ref{ssec:ssconsiderations}.

We selected from the $1,225$ lGRBs detected by \textit{Swift} up until 2020 January (inclusive) 779 lGRBs, representing 63.6 per cent of the full \textit{Swift} sample, which were localised in the RACS footprint by \textit{Swift}-XRT/UVOT to $< 15\arcsec$.
The primary goal of our search was to find and study bright late-time radio emission from this subset of well-localised GRBs.
We crossmatched our catalogue of well-localised GRBs against sources extracted from each epoch of RACS and VAST-P1 using an association radius of 15\arcsec, which is the angular resolution for the observations.
For all source associations, we used information on the positions of host galaxies and optical data (a demonstration is given in \textsection\ref{sec:results}) as well as the detectability criteria discussed in \textsection\ref{sec:detectability} to determine whether the emission was likely host or afterglow related.
For any afterglow candidate(s) found in our search, we searched archival data to determine whether any further follow-up observations were required to confirm and/or characterise the physics of the emission.

There is a possibility that some of these candidates were due to chance spatial coincidence with a background source.
We estimated the number of chance coincidence sources $\mathcal{N}_\text{cc}$ in our search with: 
\begin{equation}
    \mathcal{N}_\text{cc} \approx \mathcal{N}_\text{GRB} \times A_\text{ar} \times \rho_\text{vp}\,,
	\label{eq:coincidence}
\end{equation}
where $\mathcal{N}_\text{GRB}$ was the number of well-localised GRBs that were searched for in the RACS and VAST-P1 footprints, $A_\text{ar}$ was the area on a celestial sphere corresponding to the association radius of the targeted search, and $\rho_\text{vp}$ was the source density in the VAST-P1 data. 
Since the measured flux density for most candidates are close to the detection threshold, we used the source density at the detection threshold $\rho_\text{vp}\,(S \geq 1.50\,\text{mJy}) \approx 80$ sources per square degree along with our search parameters as inputs for Equation \ref{eq:coincidence} and estimated $\mathcal{N}_\text{cc}$ to be 3.45. 
We also estimated $\mathcal{N}_\text{cc}$ numerically by repeating the crossmatch 100 times using simulated GRB positions, with right ascension and cosine of the declination (up to declination $+41$\degr{}) generated from a random uniform distribution.
The numerical method estimates $\mathcal{N}_\text{cc} = 4.32 \pm 1.85$, which is in agreement with our analytical estimate.

\section{Results and Further Analysis}
\label{sec:results}

\begin{table*}
	\centering
	\caption{
	Radio observations of the GRB 171205A afterglow. Column 1 shows the start date of the observations or the mean epoch of the observations for those spanning more than two days; Column 2 shows the time of the observations in days post-burst; Column 3 shows the radio telescope or survey used for the observations; Column 4 shows the central frequency of the observations; Column 5 shows the flux-density measurements for detections or the 3$\sigma$-limits for non-detections; Column 6 shows references to the initial reports of the observations. Observations above the dividing line are early time observations taken within 3 months of the event and observations below the dividing line are late time observations conducted during or after the epoch of our radio afterglow search.}
	\label{tab:grb_observations}
	\begin{threeparttable}

	\begin{tabular}{lccccl} 
	    \hline
		\hline
		Date (UT) &
		\begin{tabular}[c]{@{}c@{}}$\Delta$T\\ (days)\end{tabular} & 
		\begin{tabular}[c]{@{}c@{}}Telescope\\or Survey\end{tabular} &
		\begin{tabular}[c]{@{}c@{}}$\nu$\\ (GHz)\end{tabular} &
		\begin{tabular}[c]{@{}c@{}}${S_\nu}^\dagger$\\(\si{\milli\jansky})\end{tabular} &
		Reference\\
		\hline
		
		2017 Dec 09.58 & \phantom{0}4.31 & VLA & 5.0 & $2.41 \pm 0.12$ & \citet{Urata2019,LaskarGCN22216}\\
		
		& & & 7.1 & $4.32 \pm 0.05$ & \textquotedbl\textquotedbl\\
		& & & 8.5 & $5.71 \pm 0.05$ & \textquotedbl\textquotedbl\\
		& & & 11.0\phantom{0} & $8.42 \pm 0.06$ & \textquotedbl\textquotedbl\\
		& & & 13.5\phantom{0} & $11.26 \pm 0.09$\phantom{0} & \textquotedbl\textquotedbl\\
		& & & 16.0\phantom{0} & $14.01 \pm 0.11$\phantom{0} & \textquotedbl\textquotedbl\\
		
		2017 Dec 10.07 & \phantom{0}4.76 & GMRT & 1.4 & $< 0.180$ & \citet{ChandraGCN22222}\\
		
		2017 Dec 12.00 & \phantom{0}6.70 & ATCA & 5.5 & $4.00 \pm 0.18$ & This work \\
		& & & 9.0 & $5.98 \pm 0.45$ & \textquotedbl\textquotedbl\\
		& & & 44.0\phantom{0} & $18.5 \pm 2.1$\phantom{0} & \textquotedbl\textquotedbl\\
		
		2017 Dec 12.61 & \phantom{0}7.30 & RATAN-600 & 4.7 & $<3$ & \citet{TrushkinGCN22258}\\
		& & & 8.2 & $10\pm3$ & \textquotedbl\textquotedbl\\
		2017 Dec 15.53 & 10.22 & VLBA & 4.4 & 2.3 & \cite{PTGCN22302}\\
		2017 Dec 19.07 & 13.76 & GMRT & 1.4 & $0.782\pm 0.057$ & \citet{ChandraGCN22264}\\
		2017 Dec 20.06 & 14.75 & eMERLIN & 5.1 & $7.9 \pm 0.4$ & \citet{PTGCN22350}\\
		
        2018 Jan 10.70 & 36.40 & ATCA & 4.7 & $7.48 \pm 0.47$ & This work \\
        & & & 5.2 & $7.88 \pm 0.12$ & \textquotedbl\textquotedbl\\
        & & & 5.5 & $8.07 \pm 0.13$ & \textquotedbl\textquotedbl\\
        & & & 5.8 & $8.15 \pm 0.14$ & \textquotedbl\textquotedbl\\
        & & & 6.3 & $8.47 \pm 0.14$ & \textquotedbl\textquotedbl\\
        & & & 8.2 & $8.50 \pm 0.15$ & \textquotedbl\textquotedbl\\
        & & & 8.7 & $8.61 \pm 0.16$ & \textquotedbl\textquotedbl\\
        & & & 9.0 & $8.78 \pm 0.16$ & \textquotedbl\textquotedbl\\
        & & & 9.3 & $8.69 \pm 0.17$ & \textquotedbl\textquotedbl\\
        & & & 9.8 & $8.65 \pm 0.16$ & \textquotedbl\textquotedbl\\
        & & & 44.0\phantom{0} & $15.4 \pm 1.2$\phantom{0} & \textquotedbl\textquotedbl\\

		2018 Feb 01.35 & 58.04 & VLASS & 3.0 & $5.80 \pm 0.46 $ & \citet{Lacy2020}\\
		
        2018 Feb 19.61 & 76.00 & ATCA & 4.7 & $7.28 \pm 0.27$ & This work \\
        & & & 5.2 & $7.54 \pm 0.20$ & \textquotedbl\textquotedbl\\
        & & & 5.5 & $7.70 \pm 0.21$ & \textquotedbl\textquotedbl\\
        & & & 5.8 & $8.21 \pm 0.41$ & \textquotedbl\textquotedbl\\
        & & & 6.3 & $8.02 \pm 0.32$ & \textquotedbl\textquotedbl\\
        & & & 8.2 & $8.63 \pm 0.25$ & \textquotedbl\textquotedbl\\
        & & & 8.7 & $8.73 \pm 0.28$ & \textquotedbl\textquotedbl\\
        & & & 9.0 & $8.90 \pm 0.27$ & \textquotedbl\textquotedbl\\
        & & & 9.3 & $8.94 \pm 0.29$ & \textquotedbl\textquotedbl\\
        & & & 9.8 & $9.15 \pm 0.33$ & \textquotedbl\textquotedbl\\
		\hline
		2019 Apr 30.39$^\ddagger$ & 511.08 & RACSe1 & 0.8875 & $2.51 \pm 0.76$ & This work \\ 

		2020 Mar 26.63$^\mathsection$ & 842.31 & RACSe2 & 0.8875 & $1.94 \pm 0.70$ & This work \\

		2020 Apr 12.46 & 859.15 & ATCA & 2.1 & $0.71 \pm 0.25$ & This work \\
		& & & 5.5 & $0.41 \pm 0.05$ & \textquotedbl\textquotedbl\\
		& & & 9.0 & $0.25 \pm 0.03$ & \textquotedbl\textquotedbl\\
		& & & 16.7\phantom{0} & $0.11 \pm 0.04$ & \textquotedbl\textquotedbl\\
		& & & 21.2\phantom{0} & $< 0.09$ & \textquotedbl\textquotedbl\\

		2020 May 07.26 & 883.95 & ATCA & 2.1 & $0.82 \pm 0.29$ & This work \\
		& & & 5.5 & $0.44 \pm 0.05$ & \textquotedbl\textquotedbl\\
		& & & 9.0 & $0.24 \pm 0.03$ & \textquotedbl\textquotedbl\\
		& & & 16.7\phantom{0} & $0.13 \pm 0.05$ & \textquotedbl\textquotedbl\\
		& & & 21.2\phantom{0} & $< 0.12$ & \textquotedbl\textquotedbl\\
		\hline
	\end{tabular}
	\begin{tablenotes}
      \small
      \item $^\dagger$ Uncertainties for GCN Circulars are as reported. Uncertainties for VLASS and RACS measurements consist of a statistical and systematic component ($\sim 8$ per cent for VLASS and $\sim 30$ per cent for RACS) added together in quadrature.
      \item $^\ddagger$ This is the observation of the afterglow initially detected in our search.
      \item $^\mathsection$ Some tiles of RACS were reobserved. RACSe1 refers to the initial observation and RACSe2 refers to the subsequent observation. All mentions of RACS observations in the text refer to RACSe1 unless specified otherwise.
    \end{tablenotes}
	\end{threeparttable}
\end{table*}

We found four radio source associations in our search.
Using the numerical estimate of the expected number of chance coincidence sources, we calculated the probability of all four candidates, at least one candidate and no candidates being afterglow related (rather than chance coincidence sources) to be 2.4, 65.9 and 34.1 per cent, respectively. 
Three (GRBs~080905B, 110312A and 160216A) were ruled out as imaging artefacts, host galaxy or background emission as expected from our chance coincidence calculations.
The radio source associated with GRB~171205A was the remaining afterglow candidate.
In this section, we discuss its late-time detection, our follow-up observations with the Australia Telescope Compact Array (ATCA) radio telescope, our reduction of archival early-time radio data and our compilation of multi-wavelength data on this afterglow.
More comprehensive explanations on the vetting of other candidates are provided in Appendix \ref{app:candidates}.

\subsection{Late-time emission from GRB 171205A/SN 2017iuk}
GRB 171205A was detected by the Burst Alert Telescope aboard \textit{Swift} on 2019 December 5 at 07:20:43 UT \citep{DElia2017} and an associated Type Ic-BL supernova SN 2017iuk was detected two days later \citep{dUP2017,Wang2018}.
Localised to the galaxy 2MASX J11093966$-$1235116 at $z = 0.0368$ \citep{Izzo2017}, the burst thus belongs to the subclass of local, low-luminosity GRBs with Type Ic-BL supernova association \citep{DElia2018}.
Detections of the radio afterglow prior to this work were from early-time observations as summarised in Table \ref{tab:grb_observations} (above the divider).
These include reported detections from the literature, early-time ATCA observations described in \textsection\ref{ssec:earlyatca} and a detection from an untargeted Very Large Array (VLA) Sky Survey \citep[VLASS;][]{Lacy2020} observation.
Any detections of the radio afterglow in our search and subsequent follow-up were considered late-time observations; unlike the early-time observations, these probed the behaviour of the afterglow after the synchrotron peak had passed through the radio frequencies and possibly after the blast wave had entered into the sub-relativistic phase.

In our search results, the candidate radio source associated with GRB~171205A was observed only in RACS as it was not in the VAST-P1 footprint.
In the follow-up analysis of this radio source, we used the release version of the RACS data products, with improved properties as described in \citet{McConnell2020} and Hale et al. (\emph{in prep.}), for accurate flux-density measurements, instead of the pre-release version used for our search.
The RACS tile containing the radio source was reobserved 11 months after the initial observation.
While we also provide the flux-density measurements from the reobservation in our paper, we note that at the time of our search, the tile had not been reobserved.
The flux-density measurements from both the initial and subsequent observations were low signal-to-noise detections at 6$\sigma$ to 7$\sigma$; the integrated flux-density measurements were fitting into the noise and could be unreliable as a result, so we used the peak flux-density measurements instead assuming the radio source was unresolved.

\begin{figure*}
	\includegraphics[width=0.875\linewidth,clip,trim={0cm 0cm 0cm 0cm}]{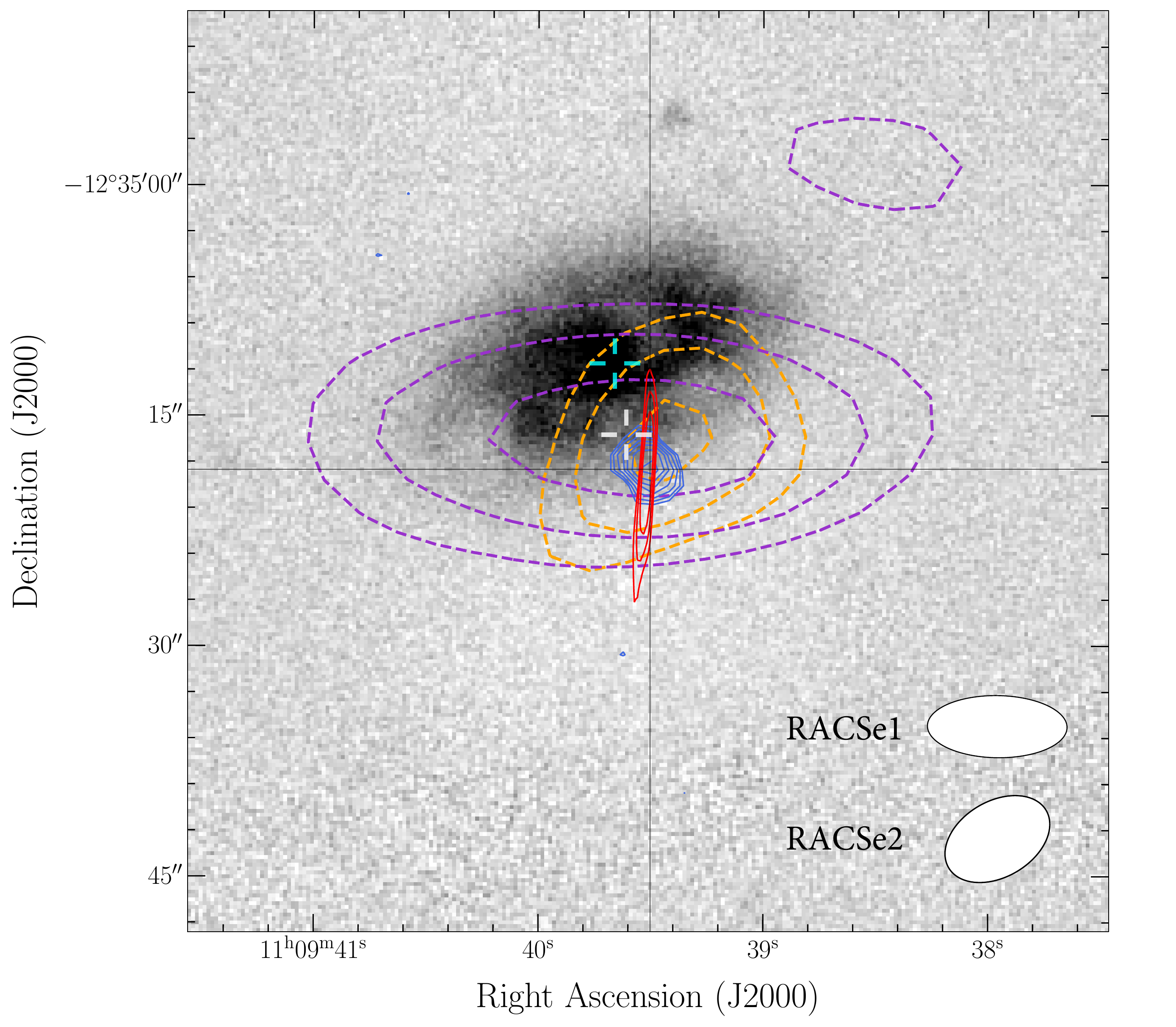}
    \caption{The Pan-STARRS $g$-band optical image of host galaxy 2MASX J11093966$-$1235116 is overlaid with radio contours from RACSe1 (purple, dashed), RACSe2 (orange, dashed), VLASS (blue, solid) and our 9.0\,GHz ATCA observations taken on 2020 May 07 (red, solid).
    The lowest contours start at the $3\sigma$-level and increase by a factor of $\sqrt{2}$ at each subsequent level.
    The PSF shapes corresponding to the RACSe1 and RACSe2 radio contours are shown on the bottom right.
    The two crosshairs represent the position of the host galaxy centre (cyan, upper) and the fitted position of the radio source in RACSe1 (white, lower).
    The size of the gap in each crosshair is approximately 1\arcsec, indicative of the astrometric uncertainty of the radio source in RACSe1.
    The image is $1\arcmin \times 1\arcmin$, centred on the observed position of the GRB optical afterglow from \citet{Izzo2017}, with North up and East to the left.}
    \label{fig:grb_panstarrs}
\end{figure*}

With only one data point at the time of the search, the source variability could not be characterised and the host-afterglow degeneracy could not be broken.
Using Equation \ref{eq:hgflux}, the range of possible SFR estimates for the host galaxy from $\sim 1\,M_\odot\ \text{yr}^{-1}$ \citep{Wang2018} to $3\pm 1 M_\odot\ \text{yr}^{-1}$ \citep{PerleyGCN22194} predicted host emission ranging from 1.2 to 4.9\,mJy; while this is consistent with the measured flux density of the source, there is considerable uncertainty on the expected flux density of the host galaxy.
To further verify whether the origin of the radio emission was related to afterglow and not host activity, we searched for evidence of positional offset between the fitted radio source position and the host galaxy.
Since this offset was unlikely to be resolvable in the RACS data, we instead searched for this in a Panoramic Survey Telescope and Rapid Response System \citep[Pan-STARRS;][]{Chambers2016} $g$-band image with sub-arcsecond resolution as shown in Figure \ref{fig:grb_panstarrs}.
This figure shows the positional offset of the radio source from the observed optical afterglow position \citep{Izzo2017} is 1\farcs2 and the positional offset from the identified host galaxy, 2MASX J20065732$-$6233465, is 4\farcs7.
The RACS tile for this source has an astrometric accuracy of $2\farcs1 \pm 3\farcs1$ in right ascension and $0\farcs3 \pm 2\farcs3$ in declination as defined by the median positional offset of sources in the tile with respect to their positions in the NRAO VLA Sky Survey \citep[NVSS;][]{Condon1998}.
These astrometric uncertainties imply that while the radio source position is more consistent with the afterglow than the host galaxy, further observations described in \textsection\ref{ssec:lateatca} were required to confirm this and to further characterise the properties of the late-time emission.

\subsection{Late-time ATCA Observations and Data Reduction}
\label{ssec:lateatca}
We observed the radio source associated with GRB 171205A twice more, on 2020 April 12 UT and on 2020 May 7 UT, with ATCA under project C3363 (PI: T. Murphy). For both epochs, the array was in the 6A configuration with a maximum baseline of 6\,km and the source was observed at multiple bands -- centred on 2.1, 5.5, 9.0, 16.7 and 21.2\,GHz -- each with a bandwidth of $2,048$\,MHz.

We reduced the visibility data using standard {\sc Miriad} procedures \citep{Sault1995}. We used a combination of manual and automatic radio-frequency interference flagging before calibration, conducted with {\sc Miriad} tasks {\sc uvflag} and {\sc pgflag}, respectively. In both epochs for all frequency bands, we used PKS B1934$-$638 to set the flux-density scale and PKS B1127$-$145 to calibrate the time-variable complex gains. We also used PKS B1934$-$638 to determine the bandpass response in both epochs for all frequency bands, except in the 16.7/21.2\,GHz bands. For these two exceptions, we used 3C 279 (B1253$-$055) as the bandpass calibrator for the first epoch and PKS B0727$-$115 for the second epoch. We determined the flux-density model for these bandpass calibrators using {\sc Miriad} task {\sc uvfmeas}. For the 2.1\,GHz band, we flagged the data from short baselines ($< 2,800$\,m) to reduce the amount of confusion from an extended, known NVSS source, J110939.0$-$123330, directly north of the afterglow candidate. This problem of flux contamination from a nearby source was only significant at 2.1\,GHz due to the lower angular resolution of the telescope in this frequency band. After calibration, we inverted and applied the multi-frequency synthesis CLEAN algorithm \citep{Hogbom1974, Clark1980, Sault1994} to the target source field using standard {\sc Miriad} tasks {\sc invert}, {\sc mfclean} and {\sc restor} to obtain our final images.

We detected a clear point source at 5.5/9\,GHz in both epochs, but it appeared slightly extended at 2.1\,GHz due to the aforementioned source confusion and at 16.7\,GHz due to low signal-to-noise.
The source was not detected at 21.2\,GHz in either epoch.
The flux densities of the radio afterglow candidate extracted from the images using {\sc Miriad} task {\sc imfit} are reported in Table \ref{tab:grb_observations} (below the divider).
For this task, we assumed a point-source model instead of a Gaussian model to avoid problems associated with fitting into the noise, especially for the 2.1- and 16.7-GHz observations.
Our detections in all frequency bands in both epochs confirmed the radio emission to be spatially consistent with the GRB/SN position, but not the host galaxy position; for example, this is clear from the radio emission at 9\,GHz from the second epoch with beam size $11\farcs67 \times 0\farcs83$ overlaid in red on top of the Pan-STARRS image in Figure \ref{fig:grb_panstarrs}.
We used a least-squares fit to constrain the spectral index $\alpha$ of the radio source in both epochs; in the first epoch, $\alpha_1 = -0.89 \pm 0.09$, and in the second epoch, $\alpha_2 = -0.99 \pm 0.15$.
With both the flux densities and spectral indices found to be consistent within their uncertainties between the two epochs, we used the data points across both epochs for a combined least-squares fit, constraining the spectral index further to $\alpha_{\text{combined}} = -0.92 \pm 0.07$.

\subsection{Early-time ATCA Observations and Data Reduction}
\label{ssec:earlyatca}
We supplemented our late-time ASKAP and ATCA data with archival early-time ATCA data to provide a more complete physical interpretation for GRB 171205A in \textsection\ref{ssec:interpretation}.
This subsection describes these early-time ATCA observations and our data reduction process.

GRB 171205A was observed three times with ATCA between 2017 December 12 and 2018 February 19 (7 to 76 days post-burst) under target-of-opportunity program CX401 (PI: M. Micha{\l}owski).
Observations were carried out using four $2,048$\,MHz bands centred on 5.5, 9, 43 and 45\,GHz for the first two epochs, while only the 5.5 and 9\,GHz bands were used for the last epoch.
We combined the 43/45\,GHz data to form a single $4,096$\,MHz band centred on 44\,GHz and split the 5.5/9\,GHz data into 512\,MHz sub-bands after calibration to image separately.

We reduced the data using {\sc Miriad}, with automated flagging applied and manual flagging where necessary, similar to our late-time ATCA data reduction in \textsection\ref{ssec:lateatca}.
We used PKS B1934$-$638 as the primary calibrator to set the flux-density scale for all observations.
We also used PKS B1934$-$638 to calibrate the bandpass at 5.5/9\,GHz, but not at 44\,GHz.
For the 44\,GHz band, we used 3C~279 (B1253-055) and PKS B1124-186 to calibrate the bandpass in the first and second epochs, respectively.
We used PKS B1127-145 and PKS B1124-186 to calibrate the time-variable complex gains at 5.5/9 and 44\,GHz, respectively.

To correct for phase errors, we performed standard self-calibration using {\sc Miriad} task {\sc gpscal}, with a single iteration used at 5.5\,GHz and two at 44\,GHz.
We detected a point source coincident with the position of GRB 171205A in each image at all three epochs. The source was marginally extended at 44\,GHz due to scatter in the model image induced by phase errors, which could not be entirely mitigated with self-calibration.
Full details of the flux-density measurements are reported in Table \ref{tab:grb_observations} (above the divider).

\subsection{Multi-wavelength Observations}
\label{ssec:mwdata}
\begin{figure*}
	\includegraphics[width=1\linewidth,clip,trim={0.3cm 0.3cm 0.3cm 0.25cm}]{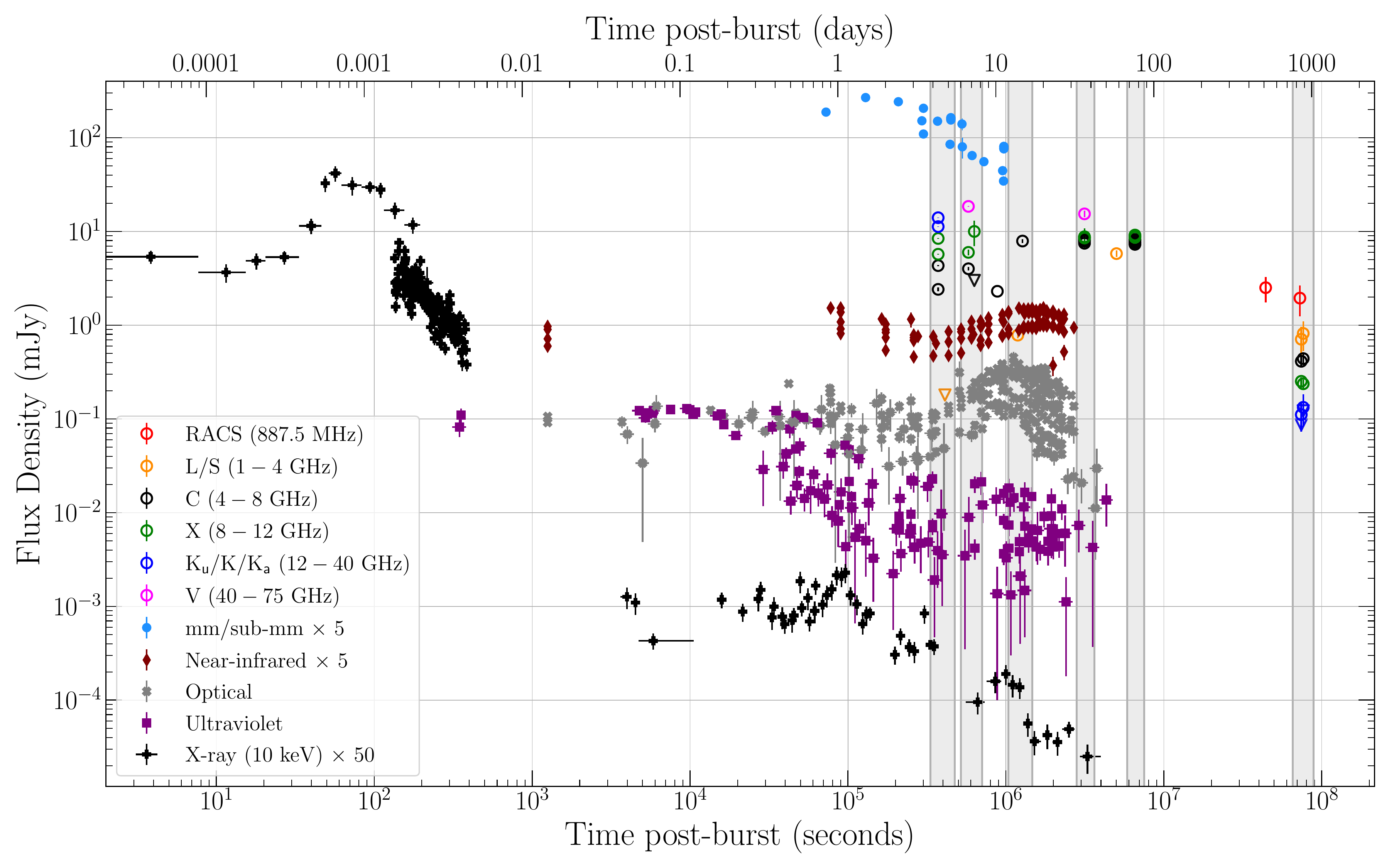}
    \caption{
    Multi-wavelength light curve for GRB 171205A/SN 2017iuk showing both the prompt, thermal and afterglow emission.
    The light curve shows data points for all observations of GRB 171205A/SN 2017iuk presented in this work and in the literature to date.
    Observations at radio frequencies probe the afterglow emission of the burst.
    They are represented by open circular markers ($3\sigma$-limits represented by open triangular markers) and are distinguished in colour by their radio band.
    The details of each radio data point are summarised in Table \ref{tab:grb_observations}.
    Observations at other wavelengths may have additional contributions from prompt and thermal supernova emission (see \textsection\ref{ssec:interpretation} text for details).
    They are represented by solid markers and are distinguished in colour by their electromagnetic spectrum classification.
    For clarity, limits are omitted and data points in the millimetre/sub-millimetre, infrared, and X--ray (10\,keV) bands have been scaled by a factor of 5, 5, and 50, respectively.
    The gray-shaded vertical stripes represent the gross epochs from which the SEDs shown in Figure~\ref{fig:sed_bb} were constructed.
    }
    \label{fig:lc}
\end{figure*}

\begin{figure*}
	\includegraphics[width=1\linewidth,clip,trim={0.3cm 0.3cm 0.3cm 0.25cm}]{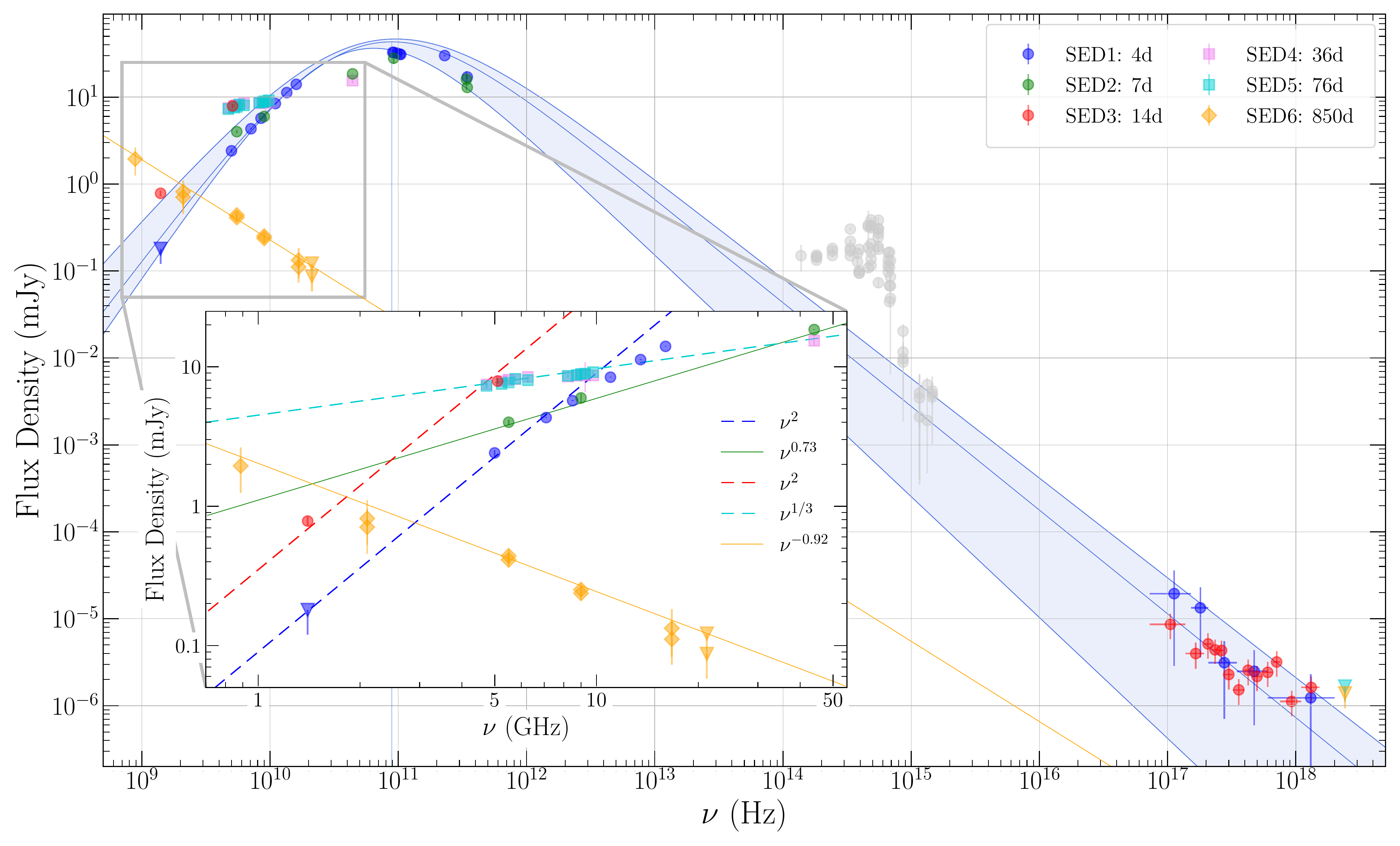}
    \caption{
    The broadband SED of the GRB 171205A afterglow taken at six epochs (approximately 4, 7, 14, 36, 76 and 850 days post-burst) from X--ray down to radio frequencies, with the inset zooming into the radio frequencies.
    Observations in the near-infrared to ultraviolet regime are shown in gray and are not included in any fits because they are contaminated by an additional thermal supernova component.
    Non-detections are given by their 3$\sigma$-limits (downward triangular markers).
    Spectral fits to data points are displayed as solid lines, while comparison lines representing typical spectral indices from segments of the standard afterglow spectrum are displayed as dashed lines.
    For SED1 (4 days post-burst), a smoothly broken power-law fit is applied (blue, with $1\sigma$-uncertainty shaded in light blue) to show the injection frequency $\nu_\text{m}$ located between the radio and millimetre/sub-millimetre frequencies.
    By analysing the evolution of the radio and broadband SED across the six epochs, we inferred the circumburst medium of the burst to be consistent with a stellar wind environment and the late-time radio emission (from RACS and ATCA observations) to be afterglow dominated (see \textsection\ref{ssec:interpretation} text for details).
    }
    \label{fig:sed_bb}
\end{figure*}

In Figure \ref{fig:lc} and Figure \ref{fig:sed_bb}, we show the multi-wavelength light curve and broadband spectral energy distribution (SED) evolution of the afterglow, respectively.
These plots bring together all the multi-wavelength data in the literature on the burst, including radio (Table~\ref{tab:grb_observations}), millimetre/sub-millimetre \citep{dUPGCN22187, PerleyGCN22252, SmithGCN22242,Urata2019}, near-infrared/optical/ultraviolet \citep{DElia2018, Izzo2019}, and X--ray \citep[flux density at $10\,$keV;][]{Evans2010}\footnote{\url{https://www.swift.ac.uk/burst_analyser/00794972/}} data points.

For the broadband SED in Figure \ref{fig:sed_bb}, we analysed the X--ray spectra using data from the \textit{Swift}-XRT, which started observing the burst 145 seconds after the trigger \citep{DElia2017}.
We extracted the XRT spectra from the public \textit{Swift} repository\footnote{\url{https://www.swift.ac.uk/user_objects/}} and analysed them with {\sc Xspec} \citep[version 12.10.1f;][]{Arnaud1996}.
We fitted the extracted spectra with an absorbed power-law model, adopting the Tubingen-Boulder ISM absorption model \citep[{\sc tbabs};][]{Wilms2000}\footnote{\url{https://heasarc.gsfc.nasa.gov/xanadu/xspec/manual/node265.html}}.
The X--ray spectrum obtained by accumulating all XRT observations starting from 4 days post-trigger (225.7\,ksec exposure time) was best fitted by an absorbed power law with equivalent hydrogen column density $N_{H}=1.22_{-0.10}^{+0.13}\times 10^{21}$\,cm$^{-2}$ (90 per cent confidence), photon index $\Gamma=1.70_{-0.32}^{+0.36}$ and de-absorbed flux integrated over the 0.3--10\,keV energy range $F=4.1\pm0.9\times 10^{-14}$\,erg\,cm$^{-2}$\,s$^{-1}$. 
This photon index, corresponding to a spectral index of $0.70_{-0.32}^{+0.36}$, is consistent with the spectral index derived from late-time radio observations in \textsection\ref{ssec:lateatca}.
In order to show the evolution of the broadband SED in Figure \ref{fig:sed_bb}, we extracted two XRT spectra at 4.3--7.3 days (SP1 -- 9\,ksec exposure time) and 7.3--14.75 days (SP2 -- 38.2\,ksec exposure time).
We fitted an absorbed power-law model with $N_{H}$ fixed to the value reported above, which is consistent with that found from the analysis of the early-time XRT data by \citet{DElia2017}.
The photon spectral indices were $\Gamma_{\rm SP1}=2.19_{-0.98}^{+1.13}$ and $\Gamma_{\rm SP2}=1.6_{-0.36}^{+1.36}$, i.e. consistent within their uncertainties.
SP1 and SP2 are represented, respectively, in Figure \ref{fig:sed_bb} with blue and red markers, corresponding to SED1 and SED3, as described below.

The data points in Figure \ref{fig:lc} (light curve) are grouped by their electromagnetic wavebands, represented by different marker shapes and colours.
The radio data points are split into further sub-bands and are distinguished from other wavelengths by open markers.
To highlight the spectral evolution of the afterglow, we grouped the observations into six gross epochs and show the SED at each of these epochs in Figure \ref{fig:sed_bb} (broadband SED).
These epochs correspond to observations taken approximately 4 days (SED1: blue), 7 days (SED2: green), 14 days (SED3: red), 36 days (SED4: pink), 76 days (SED5: cyan), and 850 days (SED6: orange) post-burst.
Data points within these epochs are not perfectly contemporaneous, but we argue this is acceptable for the subsequent analysis since the evolution of the afterglow on the timescale of days to tens of days is quite slow.
The windows for each epoch are shown by the gray vertical stripes in Figure \ref{fig:lc}.

For each epoch in the radio inset of Figure \ref{fig:sed_bb}, a spectral fit line (solid) or a comparison line corresponding to a standard afterglow spectral segment (dashed) is displayed on top of the data points.
In the wider broadband SED plot, data from the first epoch are fit to a smoothly broken power law (blue line, with $1\sigma$-uncertainty shaded in light blue):
\begin{equation}
    S_\nu = S_{\text{peak}}\bigg[\big(\frac{\nu}{\nu_\text{peak}}\big)^{-s\delta_1} + \big(\frac{\nu}{\nu_\text{peak}}\big)^{-s\delta_2}\bigg]^{-1/s}\,,
\end{equation}
where $S_{\text{peak}}$ is the flux density at the peak, $\nu_\text{peak}$ is the frequency at which the turnover occurs, $\delta_1$ is the slope of the spectral rise, $\delta_2$ is the slope of the spectral decay, and $s$ is the smoothing factor.
The Markov chain Monte Carlo fit was performed using the {\sc emcee} \citep{emcee2013} {\sc Python} implementation with uniform priors on all parameters, yielding a fit with $S_{\text{peak}} = 39.40^{+4.86}_{-5.53}$\,mJy, $\nu_\text{peak} = 54.69^{+8.93}_{-10.79}$\,GHz, $\delta_1 = 1.95^{+0.30}_{-0.20}$, $\delta_2  = -1.19^{+0.05}_{-0.20}$, and $s = 0.32^{+0.10}_{-0.08}$.
The value of the fitted peak parameters differ slightly to the true functional peak (of 43.18\,mJy located at 89.00\,GHz) due to the smoothing factor.

\section{Discussion}
\label{sec:discussion}
\subsection{GRB 171205A/SN 2017iuk Interpretation}
\label{ssec:interpretation}
The multi-wavelength light curve in Figure \ref{fig:lc} highlights many features of the GRB 171205A/SN 2017iuk system discussed in the literature.
The X--ray observations probed the prompt emission and revealed a possible thermal component in the very early phases of the X--ray light curve; the origin of which is still under debate \citep{DElia2018}. 
The near-infrared/optical/ultraviolet was dominated at early times (up to $\sim 3$ days) by a thermal component \citep{Izzo2019}. The spectroscopic studies of this emission highlighted ultra-fast expansion velocities and element abundances, which favour its origin from a hot cocoon of material produced by the jet breaking out of the stellar envelope \citep{Izzo2019}. The optical/UV light curve shows the rise of the SN 2017iuk starting $\sim 3$ days post-burst \citep{dUP2017}; spectroscopic observations confirmed this \citep{Wang2018}.
The millimetre/sub-millimetre observations probed for radio linear polarisation from the afterglow; the detection of which is still under debate \citep{Urata2019, Laskar2020}.
In this subsection, we extended the multi-wavelength analysis of this system into radio frequencies, using the data presented in Table \ref{tab:grb_observations} to conduct an empirical investigation into the afterglow properties.

We noted the radio 8--12 GHz light curve (green open markers in Figure \ref{fig:lc}) peaked around 50--70 days, i.e. after the supernova peak at $\sim 13$ days. We ascribed this radio peak as due to the evolution of the characteristic frequencies of the afterglow synchrotron spectrum across the low-frequency bands.

The radio inset of Figure \ref{fig:sed_bb} details this spectral evolution through the six epochs described in \textsection\ref{ssec:mwdata}.
In the first SED (4~days post-burst), the radio spectrum featured a steep slope (with some curvature) consistent with a spectral index $\alpha = 2$ (fitted as $1.95^{+0.30}_{-0.20}$ by the smoothly broken power law), typical for the self-absorbed segment of the afterglow spectrum.
By the second epoch (7~days post-burst), the spectrum exhibited a flatter slope fitted as $0.73 \pm 0.01$, which we interpreted to be in mid-transition towards the $\alpha = 1/3$ spectral slope, typical for the spectral segment between $\nu_\text{sa} < \nu < \nu_\text{m}$.
The slope in the third SED (14 days post-burst) remained consistent with $\alpha = 2$ as it was sampled at lower frequencies.
However, between the third and fourth epochs (14 and 36 days post-burst), a complete transition from the $\alpha = 2$ to $\alpha = 1/3$ spectral slope occurred.
In particular, this transition indicates the passage of the self-absorption frequency through $\nu = 5\,$GHz at some time between 14 and 36 days post-burst.
The flux density measurements remained similar from the fourth to fifth SED (36 and 76 days post-burst), both consistent with the typical spectral slope of $\alpha = 1/3$.

The observed SED evolution of GRB 171205A over these first five epochs provided us with clues on the properties of its circumburst medium \citep[e.g.][]{Chevalier2000}. 
The GRB jet expands into a medium whose particle density, varying as a function of distance $r$ from the progenitor, is described as $n(r) = Ar^{-k}$.
In the wind (homogeneous ISM) environment, the density drops (remains constant) with distance from the progenitor, i.e. $k=2$ ($k=0$).
While the scaling constant $A$ is just the particle density of the ISM in the homogeneous scenario, $n_0$, it is related to the mass-loss rate $\dot{M}_w$ and stellar wind velocity $v_w$ of the progenitor star by $A = \dot{M}_w/4\pi v_w = 5 \times 10^{11} A_*\,\text{g}\,\text{cm}^{-1}$ in the wind scenario.
Here, $A_*$ is a normalisation of $A$, such that $A_* = 1$ represents a typical Wolf-Rayet star with mass-loss rate $\dot{M}_w = 10^{-5}\,M_\odot\,\text{yr}^{-1}$ and stellar wind velocity $v_w = 1,000\,\text{km}\,\text{s}^{-1}$.

We argue the GRB 171205A/SN 2017iuk event occurred in a stellar wind environment induced by the progenitor prior to its core-collapse (see also \citealt{Suzuki2019}).
The radio spectral evolution in the first five epochs, in particular, the emergence of the $\alpha = 1/3$ slope in the fourth and fifth SED, was due to the passage of the self-absorption frequency, $\nu_\text{sa}$, across the gigahertz frequency range.
According to standard afterglow theory (e.g. \citetalias{Granot2002}), $\nu_\text{sa}$ only evolves with time, shifting towards lower frequencies, in the wind scenario.
The flux-density levels in the fourth and fifth SED, sampled after the passage of $\nu_\text{sa}$ through the band, were also very similar; this is consistent with the wind scenario, where the flux density is predicted to remain constant in the $\nu_\text{sa} < \nu < \nu_\text{m}$ spectral segment \citepalias{Granot2002}.
Our conclusion that the afterglow is produced by the relativistic outflow decelerating in a circumburst medium with a wind density profile is in agreement with the conclusion obtained by the independent analysis of \citet{Maity2021} using observations from the upgraded Giant Metrewave Radio Telescope.

The very late-time SED (850 days post-burst) enabled us to constrain the slope of the electron spectrum $p$, where the distribution of electron energies is $N(E) \propto E^{-p}$.
According to \citet{Maity2021}, even at $\sim 1,000$ days post-burst, the blast wave had not transitioned to the non-relativistic (Newtonian) regime.
A spectral inversion occurred between the fifth and sixth SED (76 and 850 days post-burst), where the spectral index transitioned from $\alpha = 1/3$ to $\alpha = -0.92$.
This indicates that $\nu_\text{m}$ had moved from between the millimetre and radio frequencies at 4 days post-burst and passed through the entire radio frequency band shown in the SED at some time between 76 and 850 days post-burst; by 850 days post-burst, $\nu_\text{m}$ was located below the $887.5\,$MHz sampling band.
This very late-time SED, as sampled by the radio data, is therefore located in the spectral segment $\nu_\text{m} < \nu < \nu_\text{c}$, where $\nu_\text{c}$ is the frequency of the cooling break, with characteristic spectral index $\alpha = (1-p)/2$.
Comparing this with the fitted spectral index of this SED attained in \textsection\ref{ssec:lateatca}, $\alpha = -0.92 \pm 0.07$, we constrained the slope of the electron spectrum to $p = 2.84$.
While this value deviates from the typical $p \sim 2.2$ expected from Fermi acceleration processes, it is consistent to within one standard deviation of the distribution of $p$-values obtained from afterglow modelling \citep{Ryan2015, Wang2015, Ghisellini2009}.
Moreover, particle-in-cell simulations \citep[e.g.][]{Sironi2011} have shown that possible variations in the magnetic structure at the shock front could produce different values of $p$.

The fit of SED1 (blue points in Figure \ref{fig:sed_bb}) with a single spectral component between the peak in the radio and the X--ray data points suggests that already at these early epochs the cooling frequency $\nu_\text{c}$ lies above the X--ray energy range.
This is consistent with the similar spectral slopes found in the two X--ray spectra we analysed since $\nu_\text{c}$ should increase with time in the wind scenario.

The broadband SED corresponding to SED1--3 in Figure \ref{fig:sed_bb} shows that the radio/millimetre (above 89\,GHz) through X--ray spectrum is consistent with a single non-thermal component, while the near-infrared/optical data points (gray markers) contaminated by thermal supernova emission are located above the corresponding empirical fit (blue line with shaded region). Moreover, the extrapolation of the very late-time $\alpha = -0.92$ spectral slope from the radio frequencies through to the X--rays is consistent with the very late-time upper limits derived from XRT observations. Along with the consistency of the X--ray spectral indices presented in \textsection\ref{ssec:mwdata} with the very late-time radio spectral index, these suggest the emission we sampled in the radio band starting from 4 days post-burst was dominated by the afterglow component.

As the radio SEDs are dominated by the afterglow component of the burst event, we used these observables to provide reasonable estimates for the microphysical shock parameters of the blast wave, including the normalised wind-density parameter $A_*$, the fractional energy in electrons $\epsilon_e$, and the fractional energy in the magnetic field $\epsilon_B$.
These observables are related to the microphysical shock parameters by a set of analytical relations detailed in \citet{Panaitescu2000}. For a stratified wind medium, the self-absorption frequency is:
\begin{equation}
    \nu_\text{sa} = 3.7 \times 10^9\,E^{-2/5}_{53}\,A^{6/5}_{*}\,\epsilon^{-1}_{e,-1}\,\epsilon^{1/5}_{B,-2}\,T^{-3/5}_{d}\,\text{Hz}\,,
    \label{eq:nu_sa}
\end{equation}
where $E = 10^{53} E_{53}$\,erg is the kinetic energy of the blast wave and $T_d$ is the time (in days) since the burst explosion\footnote{We note that times and frequencies should be converted from the source to the observer frame accounting for the source redshift. However, in this case, the $z=0.0368$ would introduce corrections of the order unity.}.
The flux density at the self-absorption frequency, $S_{\nu,\,\text{sa}}$, is then related to $\nu_\text{sa}$ at time $T_d$ by the relation:
\begin{equation}
    S_{\nu,\text{sa}} = 0.07 \,D^{-2}_{L,28}\,E_{53}\,A^{-1}_{*}\,\epsilon_{e,-1}\,\nu^2_{\text{sa},9.7}\,T^{1}_{d,-1}\,\text{mJy}\,,
    \label{eq:S_nu_sa}
\end{equation}
where $D_{L} = ~10^{28}D_{L,28}$\,cm is the luminosity distance of the source.
A similar relation to Equation \ref{eq:nu_sa} providing the location of the injection frequency, $\nu_\text{m}$, at time $T_d$ is given by:
\begin{equation}
    \nu_\text{m} = 1.9 \times 10^{13}\,E^{1/2}_{53}\,\epsilon^{2}_{e,-1}\,\epsilon^{1/2}_{B,-2}\,T^{-3/2}_{d}\,\text{Hz}\,.
    \label{eq:nu_m}
\end{equation}

To estimate the microphysical shock parameters using our observables and these relations, we fixed the luminosity distance parameter to $D_{L}~=~5.17 \times 10^{26}$\,cm corresponding to the afterglow redshift of $z=0.0368$ \citep{Izzo2017} and fixed the energy parameter to $E_{53} = 0.00218$ corresponding to the approximation of the blast-wave kinetic energy as ten times the energy released in the prompt emission $E_\text{prompt} = 2.18 \times 10^{49}$\,erg \citep{DElia2018}.

In Equations \ref{eq:nu_sa} and \ref{eq:S_nu_sa}, we estimated the time $T_d$ and flux density $S_{\nu,\,\text{sa}}$ at which the self-absorption frequency $\nu_\text{sa}$ passed 5\,GHz using SED3 and SED4 (14 and 36 days post-burst) in Figure \ref{fig:sed_bb}.
In the wind scenario, the flux density is expected to rise linearly with $t$ in the self-absorbed spectral segment and stay constant in the spectral segment between $\nu_\text{sa}$ and $\nu_\text{m}$.
Given this, we argue, with the 5.1\,GHz data point shown on SED3 lying on the $\alpha=1/3$ line alongside SED4 data points, $\nu_\text{sa}$ was likely already at 5\,GHz around 14 days post-burst.
We thus used the self-absorption break at $\nu_\text{sa} = 5$\,GHz with $S_{\nu,\text{sa}} = 7.9$\,mJy and $T_d = 14$ days as our observables for Equations \ref{eq:nu_sa} and \ref{eq:S_nu_sa}.
These numbers are only estimates since we do not see a clear self-absorption break in any SED; $\nu_\text{sa}$ may have crossed 5\,GHz at any time between 14 and 36 days.
For Equation \ref{eq:nu_m}, while none of the radio SEDs in Figure~\ref{fig:sed_bb} show a spectral break corresponding to the injection frequency $\nu_\text{m}$, SED6 (850 days post-burst) provides an observable boundary condition; it suggests $\nu_\text{m}$ at $T_d = 850$ days post-burst lies below $887.5$\,MHz.

To estimate $A_*$, $\epsilon_e$ and $\epsilon_B$, we solved the set of non-linear Equations \ref{eq:nu_sa}--\ref{eq:nu_m} using a non-linear least-squares optimisation routine in the {\sc SciPy} \citep{SciPy2020} {\sc Python} library.
We imposed the following constraints on the free parameters: $A_* > 0$, $0 < \epsilon_e,\,\epsilon_B < 1$ and $0 < \nu_\text{m} < 887.5$\,MHz; the last constraint implemented the boundary condition associated with SED6 and Equation~\ref{eq:nu_m}.
This yields us the following estimates for the microphysical shock parameters: $A_* = 3$, $\epsilon_e = 0.3$ and $\epsilon_B = 0.0002$, where the boundary-defining parameter takes the value $\nu_\text{m} = 46$\,MHz for this solution.
The low value of $\epsilon_B$ we derived is consistent with the wide distributions of values estimated from the modelling of GRB afterglows (e.g. $10^{-10}$ to $10^{-3}$ in \citealt{BD2014}, $10^{-6}$ to $10^{-3}$ in \citealt{Santana2014}).
These low values can explain the properties of the afterglow emission at very high energies as detected by \textit{Fermi}-LAT \citep{Beniamini2015} and have been interpreted to be due to the possible decay of the magnetic field in the downstream region behind the external shock \citep{Lemoine2013}.
While we have shown plausible estimates of the microphysical parameters can be derived from our data, these estimates are very sensitive to our assumptions; more accurate estimates would require full afterglow modelling across relativistic and sub-relativistic regimes, which is beyond the scope of this paper.

\subsection{Search Strategy Considerations}
\label{ssec:ssconsiderations}
In this work, we limited our search to well-localised bursts to reduce the number of false, or coincident, associations.
Since this localisation is often achieved by \textit{Swift}-XRT/UVOT, we introduced a bias towards bursts with detectable afterglows at X--ray, UV and optical wavelengths. For example, bursts with detectable afterglows are more likely to be associated with certain physical properties, such as a higher circumburst density \citep[e.g.][]{Piran2004}.
If we removed this localisation requirement and searched for radio afterglow candidates within the GRB error radius, we would be able to provide follow-up observations for approximately three times as many, or $\sim 2,000$ more, GRBs.

In this scenario, the problem of false associations becomes significant.
Using a typical error radius of $5\arcmin$ in Equation \ref{eq:coincidence}, we find that the number of expected false associations would be $1,370$.
With poor localisation and large number of false associations, our method for distinguishing afterglow and host-related emission by analysing positional offsets in optical images would not be very effective.
Instead, a search without the localisation criterion would need an additional variability requirement to reduce the number of false associations (the source density used in Equation \ref{eq:coincidence} would then be the density of variable sources in VAST-P1).
Since the VAST-P1 footprint is the only region with multiple epochs in our search, a search requiring variability analysis would reduce our sky coverage by a factor of $\sim 7$ (when compared with including RACS), and correspondingly, the expected afterglow detection rates estimated in \textsection\ref{sec:rates} would also decrease by a factor of $\sim 7$.
When a deep, single epoch sky map from EMU (see \textsection\ref{ssec:future_yields}) is completed, it can be used as a reference epoch for this type of time-domain analysis; this will reduce the false association rate for searches in future surveys, regardless of whether these surveys are multi-epoch or not.

Due to these considerations, there were significant challenges in using previous generations of wide coverage radio surveys, in particular, NVSS and SUMSS, to conduct a similar search.
These surveys conducted in the pre-\textit{Swift} era would have significantly fewer well-localised bursts to follow-up and the lack of multi-epoch data hindered the ability to conduct further variability analysis for these bursts with larger error radius.

Given our choice to limit our search to well-localised bursts, we also did not consider any bursts occurring prior to 2004, the launch year of \textit{Swift}.
Due to this constraint, our search was unable to test late-time rebrightening theories for GRB/SN systems since this phenomenon is predicted to occur on the timescales of a few decades and usually at sub-mJy flux-density levels \citep[e.g.][]{Peters2019}.
Still, our search was sensitive to local, radio-bright GRB/SN afterglows as demonstrated by the detection of late-time emission from the GRB 171205A/SN 2017iuk system.
This suggests orphan afterglow searches using the current generation of wide coverage radio surveys, such as RACS and VAST-P1, may also be sensitive enough to detect off-axis emission from these systems, providing an alternative to orphan afterglow searches guided by supernova association \citep[e.g. off-axis GRB detection via association with Type Ic-BL SN 2020bvc;][]{Izzo2020}.

\subsection{GRB Yields in Comparison Radio Surveys}
\label{ssec:future_yields}
Our ASKAP search for radio afterglows probed for radio emission from 587 GRBs that were not previously observed by the Arcminute Microkelvin Imager \citep[AMI;][]{Anderson2018}, ATCA, or VLA, and resulted in one detection.
With the sensitivity improvements expected in future radio surveys, for example, the SKA-Shallow survey targeting a 5$\sigma$-detection threshold of 1\,\si{\micro\jansky\,beam^{-1}}, similar searches will be able to study greater classes of GRB events, including sGRBs, useful for identifying electromagnetic counterparts to gravitational wave events, and high-redshift GRBs, useful for probing Population-III stars.
These expected improvements to sensitivity in such searches will enable sufficient detections for population analyses of radio afterglows to be conducted, providing useful experiments to better understand the nature of radio-quiet and -loud GRBs, for example.
In this subsection we compare the GRB yields of our search using RACS and VAST-P1 to various surveys, including the EMU Pilot Survey, commensal ASKAP surveys and the SKA-Shallow survey.

Our search covered $\sim 40,000$ square degrees and the detection of one low-luminosity GRB afterglow above our 5$\sigma$-sensitivity threshold of $\sim 1.50$\,\si{\milli\jansky\,beam^{-1}} implies a transient surface density $\rho \gtrsim 1 \times 10^{-4}$\,deg$^{-2}$.
This is consistent with the gigahertz frequency transient surface density prediction for low-luminosity GRBs in Figure 3 of \citet{Metzger2015} calculated to be $\rho \approx 5 \times 10^{-4}$\,deg$^{-2}$.
Following our detection of the GRB 171205A radio afterglow up to 842 days post-burst in RACS, we extended the time baseline for our rates analysis in \textsection\ref{sec:rates} from 2011 to 2020.
We searched online databases\footnote{Jochen Greiner database: \href{https://www.mpe.mpg.de/~jcg/grbgen.html}{\texttt{www.mpe.mpg.de/\char`\~jcg/grbgen.html}}}\footnote{\textit{Swift} database: \href{https://swift.gsfc.nasa.gov/archive/grb_table.html}{\texttt{swift.gsfc.nasa.gov/archive/grb\_table.html}}} for radio afterglow detections up to 2020 January and manually searched the literature for radio flux-density measurements of each detection.
We found that aside from GRB 171205A, no lGRB radio afterglows reached a peak flux density above the 5$\sigma$-sensitivity threshold of $\sim 1.50$\,\si{\milli\jansky\,beam^{-1}} at 887.5\,MHz.
However, the radio afterglow from Very High Energy ($E > 100$\,GeV) lGRB 190829A, occurring in the VAST-P1 footprint two days after our first epoch of observations, may have been marginally detectable, i.e. at $4\sigma$-significance, with more favourable timing, having peaked at $1.52 \pm 0.22$\,\si{\milli\jansky\,beam^{-1}} at an observing frequency of 1.3\,GHz \citep{Rhodes2020}.
The new information improves the probability of detecting a radio afterglow calculated in \textsection\ref{sec:rates} to 13.3 per cent (11.3 per cent), assuming the $\alpha=1/3\ (2)$ spectral scaling scenario.
While this probability suggests the detection of GRB 171205A afterglow in our search was unlikely, its detection could be attributed to the extreme nature of the event, being the second brightest afterglow detected to date at radio frequencies.

Surveys reaching lower flux-density thresholds are required to probe for classes of GRBs beyond the subset of local, radio-bright GRB/SN systems.
We consider the expected GRB yield of some of these surveys here, assuming a more conventional 1.4-GHz observing frequency, spectral scaling with $\alpha=1/3$ where necessary and an optimistic one month afterglow duration \citep[the typical duration differs significantly from the very local GRB/SN systems considered in \textsection\ref{sec:rates} and would vary depending on the flux-density threshold; see, for example, \citetalias{Chandra2012} and][]{Ghirlanda2013}.
The EMU Pilot Survey, conducted with the same amount of observation time as VAST-P1, reached a $5\sigma$-sensitivity threshold of 100\,\si{\micro\jansky\,beam^{-1}}, but covers $\sim 10$ per cent ($\sim 2$ per cent) of the VAST-P1 (RACS) sky regions as shown in Figure \ref{fig:vastskymap}.
This limited sky coverage produces a worse probability for radio afterglow detection of 0.3 per cent, after repeating the analysis in \textsection\ref{sec:rates} for the EMU Pilot Survey.
Unlike the subset of GRBs with supernovae association previously considered, radio follow-up of typical lGRBs at more cosmological distances is more sparse (only 18 per cent of lGRBs receive radio follow-up), meaning the use of the \citepalias{Chandra2012} sample alone underestimates the detection probability for the EMU Pilot Survey; yet, accounting for this only boosts the probability of detection up to 1.6 per cent.

Aside from the consideration of sky coverage, analysing positional offsets in optical images as demonstrated in this work will not be enough to distinguish afterglow from host emission for the EMU Pilot Survey and other deeper surveys.
As shown in Figure \ref{fig:onaxisrates}, deeper surveys will detect afterglows up to a higher redshift, but the host-afterglow offset at higher redshifts will be too small.
For example, GRB 171205A at $z=0.0368$ had a noticeable angular offset of 5\farcs9 from its host, translating to a physical offset of $\sim 4.4$\,kpc; but if it had a typical (median) lGRB physical offset of $\sim 1.3$\,kpc instead, the angular offset would be $<2$\arcsec.
Variability analysis over multiple epochs would be required and this is not possible with the single epoch EMU Pilot Survey.

The full ASKAP surveys (e.g. EMU and VAST) will be conducted with a commensal approach; depending on the cadence, multi-epoch deep surveys with large sky coverage may be available.
For example, the full EMU survey aims to cover the entire RACS footprint at the EMU Pilot Survey sensitivity, requiring $\sim 10$\,hr of integration per tile.
In the extreme cases, the 10\,hr could be observed through 10\,min pointings with 60 visits per tile over 5 years (i.e. 1 visit per month) or through two 5\,hr pointings per tile separated by over a year; a hybrid between these two extremes is likely more realistic.
The former will reach a $5\sigma$-sensitivity threshold similar to our RACS/VAST-P1 search and has a 47.8 per cent probability (based of a similar analysis to that in \textsection\ref{sec:rates}) of detecting a radio afterglow from a known local radio-bright GRB/SN system, while the latter will reach a $5\sigma$-sensitivity threshold of $\sim 150$\,\si{\micro\jansky\,beam^{-1}} and has a 62.4 per cent (16.2 per cent if we do not account for GRBs that do not receive radio follow-up) probability of detecting a radio afterglow from a broader class of known on-axis GRBs.

These ASKAP surveys, which are expected to yield a number of radio afterglow detections of order unity, are precursors to more potent SKA-era surveys.
The proposed SKA-Shallow survey, for example, covering half the sky to sub-\si{\micro\jansky} sensitivity with daily cadence would provide radio follow-up to all bursts in the half-sky and deeper non-detection limits than any other radio telescope to date.
While the ASKAP surveys used in our work still do not reach the flux-density sensitivity required for placing stringent tests on radio rebrightening theories and conjectures regarding distinct radio-quiet/loud GRB populations, applying a similar search methodology and analysis to SKA-era surveys will enable these experiments as well as further insights into the general radio afterglow population to be realised.

\section{Conclusions}
\label{sec:conclusions}
We have conducted a search for radio afterglows from 779 well-localised lGRBs occurring after 2004 by crossmatching their positions with radio sources in the RACS and VAST-P1 data.
Analysing the positional offset of the radio source and host galaxy in optical images was the primary method we used to distinguish afterglow from background or host related emission for matched sources.
Our search produced four candidates, three of which were ruled out as background source or host-galaxy related, while one was confirmed as a radio afterglow from GRB 171205A, a low-luminosity lGRB associated with SN 2017iuk.

We conducted further late-time observations of the GRB~171205A radio afterglow with ATCA at 859 and 884 days post-burst.
Both the flux densities and spectral indices were measured to be consistent within their uncertainties between the two epochs.
Our late-time radio observations provided a robust measurement of the electron spectral index and also provided the bounded constraint on the injection frequency (see Equation \ref{eq:nu_m}) required to solve for the other microphysical parameters.
Using this late-time data alongside archival data from early-time radio observations, we showed the progenitor of this GRB-SN system originated in a stellar wind environment with microphysical shock parameters of the burst estimated as $p = 2.84$, $A_* = 3$, $\epsilon_e = 0.3$ and $\epsilon_B = 0.0002$.

While we estimated a probability of $< 10$ per cent for detecting any radio afterglows in our search, we were able to probe for a subset of radio-bright lGRBs in the local Universe at $z<0.2$ (or $D_L < 1$\,Gpc) that were often accompanied by supernova detections.
In order to probe for other GRB classes through deeper surveys or to overcome biases from choosing to only follow-up well-localised lGRBs, multiple epochs of data will be required in addition to the analysis of positional offsets in optical images.
The integration time and cadence of these epochs in full commensal ASKAP surveys will determine the classes of GRBs similar searches would be sensitive to in the future.
Our method of analysing positional offsets in optical images will also be applied in ongoing searches in our RACS and VAST-P1 data for explosive transients expected to be offset from a host, including orphan afterglows, magnetars, fast radio burst afterglows, and more.

\section*{Acknowledgements}
We thank the anonymous referee and Phil Edwards for useful comments that improved the quality of this manuscript.
JL, DD and JP are supported by Australian Government Research Training Program Scholarships.
TM acknowledges the support of the Australian Research Council through grants FT150100099 and DP190100561.
DK is supported by National Science Foundation grant AST-1816492.
Parts of this research were conducted by the Australian Research Council Centre of Excellence for Gravitational Wave Discovery (OzGrav), through project number CE170100004.

The Australian Square Kilometre Array Pathfinder is part of the Australia Telescope National Facility which is managed by CSIRO. 
Operation of ASKAP is funded by the Australian Government with support from the National Collaborative Research Infrastructure Strategy. ASKAP uses the resources of the Pawsey Supercomputing Centre. Establishment of ASKAP, the Murchison Radio-astronomy Observatory and the Pawsey Supercomputing Centre are initiatives of the Australian Government, with support from the Government of Western Australia and the Science and Industry Endowment Fund. 
We acknowledge the Wajarri Yamatji as the traditional owners of the Murchison Radio-astronomy Observatory site.
The Australia Telescope Compact Array is part of the Australia Telescope National Facility which is funded by the Australian Government for operation as a National Facility managed by CSIRO.
We acknowledge the Gomeroi people as the traditional owners of the Paul Wild Observatory site.

The Pan-STARRS1 Surveys (PS1) and the PS1 public science archive have been made possible through contributions by the Institute for Astronomy, the University of Hawaii, the Pan-STARRS Project Office, the Max-Planck Society and its participating institutes, the Max Planck Institute for Astronomy, Heidelberg and the Max Planck Institute for Extraterrestrial Physics, Garching, The Johns Hopkins University, Durham University, the University of Edinburgh, the Queen's University Belfast, the Harvard-Smithsonian Center for Astrophysics, the Las Cumbres Observatory Global Telescope Network Incorporated, the National Central University of Taiwan, the Space Telescope Science Institute, the National Aeronautics and Space Administration under Grant No. NNX08AR22G issued through the Planetary Science Division of the NASA Science Mission Directorate, the National Science Foundation Grant No. AST-1238877, the University of Maryland, Eotvos Lorand University (ELTE), the Los Alamos National Laboratory, and the Gordon and Betty Moore Foundation.
The national facility capability for SkyMapper has been funded through ARC LIEF grant LE130100104 from the Australian Research Council, awarded to the University of Sydney, the Australian National University, Swinburne University of Technology, the University of Queensland, the University of Western Australia, the University of Melbourne, Curtin University of Technology, Monash University and the Australian Astronomical Observatory. SkyMapper is owned and operated by The Australian National University's Research School of Astronomy and Astrophysics.
This work made use of data supplied by the UK Swift Science Data Centre at the University of Leicester.

\textit{Software:} {\sc Astropy} \citep{Astropy2013, Astropy2018}, {\sc emcee} \citep{emcee2013}, {\sc matplotlib} \citep{Hunter2007}, {\sc NumPy} \citep{Harris2020}, {\sc pandas} \citep{McKinney2010}, {\sc SciPy} \citep{SciPy2020}, {\sc ASKAPsoft} \citep{Cornwell2011, Guzman2019}, {\sc Selavy} \citep{Whiting2012}, {\sc Miriad} \citep{Sault1995}, and {\sc Xspec} \citep{Arnaud1996}.

\section*{Data Availability}
The ASKAP data used in this paper (RACS and VAST-P1) can be accessed through the CSIRO ASKAP Science Data Archive (CASDA\footnote{\url{https://data.csiro.au/dap/public/casda/casdaSearch.zul}}) under project codes AS110 and AS107.
The ATCA data used in this paper can be accessed through the Australia Telescope Online Archive (ATOA\footnote{\url{https://atoa.atnf.csiro.au/query.jsp}}) under project codes CX401 and C3363.
Other auxiliary datasets can be made available upon request via email to the corresponding author.


\bibliographystyle{mnras}
\bibliography{bibliography}


\newpage
\appendix
\section{Ruled-out Afterglow Candidates}
\label{app:candidates}
\begin{figure*}
	\includegraphics[width=0.85\linewidth]{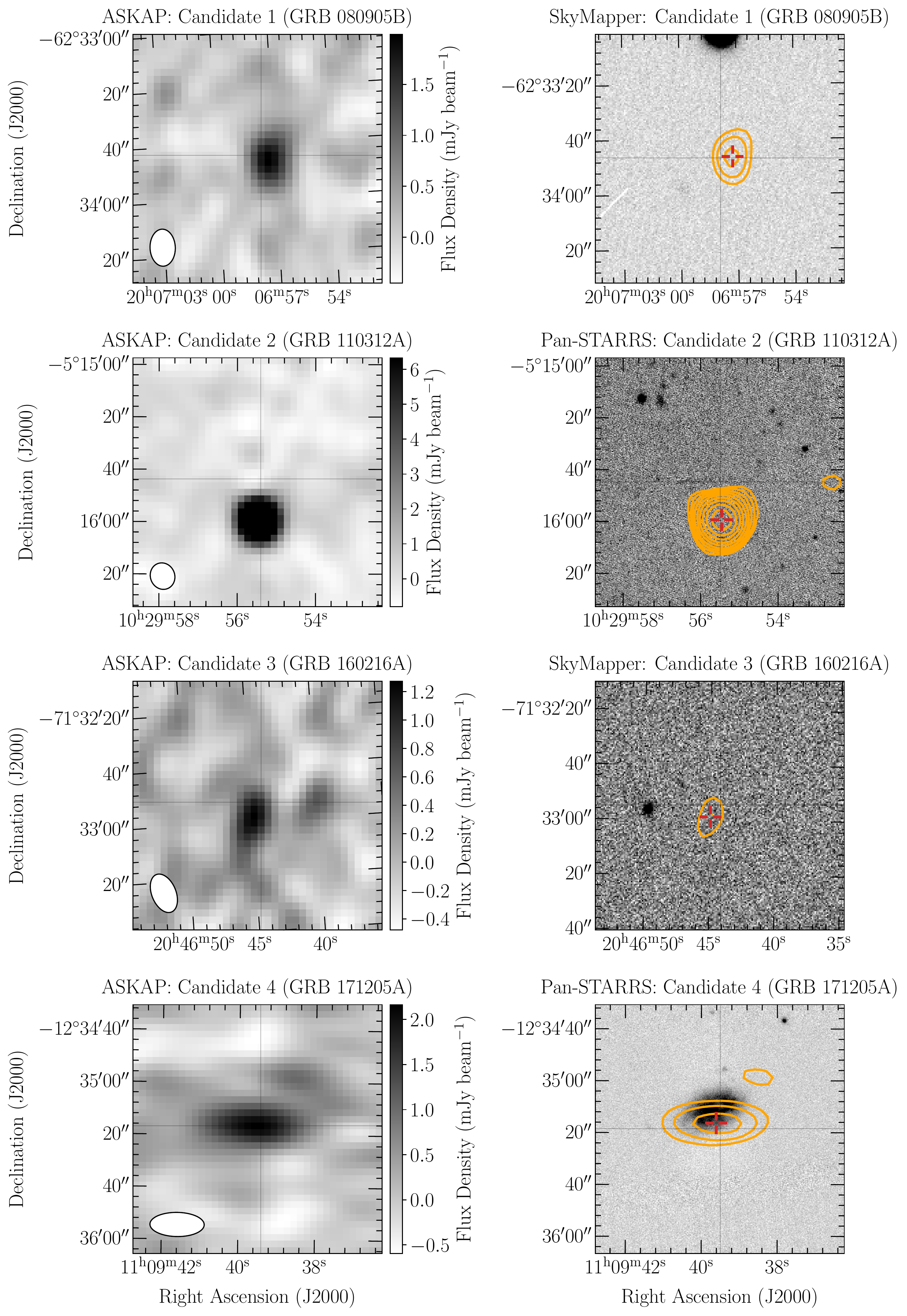}
    \caption{
    The ASKAP image (left) and optical image (right) for each afterglow candidate found in our search.
    All ASKAP images are from RACS, with the exception of the Candidate 2 (GRB~110312A) ASKAP image taken from epoch 8 of VAST-P1.
    The optical images shown are $g$-band images from either SkyMapper \citep[][Candidates 1 and 3]{Onken2019} or Pan-STARRS (Candidates 2 and 4) with radio contours from their corresponding ASKAP images overlaid (orange).
    The lowest contours start at the $3\sigma$-level and increase by a factor of $\sqrt{2}$ at each subsequent level.
    The red crosshair represents the fitted radio position of the afterglow candidate.
    All images are $1\farcm5 \times 1\farcm5$, centred on their corresponding \textit{Swift}-XRT/UVOT localised afterglow positions, with typical localisation errors of a few arcseconds.
    North is up and East is to the left.
    }
    \label{fig:candidates}
\end{figure*}

In \textsection\ref{sec:results}, we ruled out three of the four afterglow candidates in the vetting process. For each candidate, we provide an explanation and show in Figure \ref{fig:candidates} their corresponding radio and optical images.

\textbf{Candidate 1 (GRB~080905B)} -- This matched radio source found in RACS is likely to be related to the host galaxy, 2MASX J20065732$-$6233465.
The fitted radio source position is more spatially consistent with the host position ($0\farcs7$ offset) than with the GRB position ($4\farcs5$ offset).

\textbf{Candidate 2 (GRB~110312A)} -- This matched radio source is the only candidate located within the VAST-P1 footprint and it was detected in all epochs of our search.
The emission is either host galaxy related or from a coincident background source since the radio source is not spatially consistent with the GRB position ($14\farcs3$ offset).

\textbf{Candidate 3 (GRB~160216A)} -- This matched radio source found in RACS is not spatially consistent with the GRB position ($5\farcs7$ offset) and is likely to be either host galaxy related or from a coincident background source.
It is also possible this is an imaging artefact caused by a local noise peak close to the detection threshold.

\textbf{Candidate 4 (GRB~171205A)} -- This matched radio source found in RACS is a confirmed afterglow candidate.
The details are discussed in \textsection\ref{sec:results}.

\bsp	
\label{lastpage}
\end{document}